\documentclass[11pt]{article}
\usepackage{jcappub}
\usepackage{bm}
\usepackage{dsfont}
\usepackage{color}
\usepackage{array}
\usepackage{graphicx}
\usepackage{soul}
\usepackage{multirow}
\usepackage{multicol}
\usepackage{float}
\usepackage{hhline}
\usepackage[dvipsnames]{xcolor}
\usepackage[normalem]{ulem}
\usepackage{mathrsfs}
\usepackage{wasysym}
\usepackage[mathscr]{euscript}
\newcolumntype{P}[1]{>{\centering\arraybackslash}p{#1}}
\newcolumntype{M}[1]{>{\centering\arraybackslash}m{#1}}

\newcommand{\be}{\begin{equation}}
\newcommand{\ee}{\end{equation}}

\newcommand{\een}{\end{subequations}}
\newcommand{\ben}{\begin{subequations}}

\newcommand{\lsim}{\mathrel{\mathop{\kern 0pt \rlap
      {\raise.2ex\hbox{$<$}}}\lower.9ex\hbox{\kern-.190em $ \sim$}}}
\newcommand{\gsim}{\mathrel{\mathop{\kern 0pt
      \rlap{\raise.2ex\hbox{$>$}}}\lower.9ex\hbox{\kern-.190em $\sim$}}}

\newcommand{\CO}{\mathcal{O}}

\title{Complementarity of experiments in probing the non-relativistic effective theory of dark matter-nucleon interactions}

\author[a,b]{Anja Brenner,}
\author[a,b]{Gonzalo Herrera,}
\author[b]{Alejandro Ibarra,}
\author[c]{Sunghyun Kang,}
\author[c]{Stefano Scopel,}
\author[b]{Gaurav Tomar}
\emailAdd{anja.brenner@tum.de}
\emailAdd{gonzalo.herrera@tum.de}
\emailAdd{ibarra@tum.de}
\emailAdd{francis735@naver.com}
\emailAdd{scopel@sogang.ac.kr}
\emailAdd{physics.tomar@tum.de}
	\affiliation[a]{\normalsize\textit{Max-Planck-Institut f\"ur Physik (Werner-Heisenberg-Institut),\protect\\ F\"ohringer Ring 6, 80805 M\"unchen, Germany}}
	\affiliation[b]{\normalsize\textit{Physik-Department, Technische Universit\"at M\"unchen, \protect\\James-Franck-Stra\ss{}e, 85748 Garching, Germany}}
\affiliation[c]{\normalsize\textit{Department of Physics, Sogang University, Seoul 121-742, South Korea}}

\abstract{The non-relativistic effective theory of dark matter-nucleon interactions depends on 28 coupling strengths for dark matter spin up to 1/2. Due to the vast parameter space of the effective theory, most experiments searching for dark matter interpret the results assuming that only one of the coupling strengths is non-zero. On the other hand, dark matter models generically lead in the non-relativistic limit to several interactions which interfere with one another, therefore the published limits cannot be straightforwardly applied to model predictions. We present a method to determine a rigorous upper limit on the dark matter-nucleon interaction strength including all possible interferences among operators. We illustrate the method to derive model independent upper limits on the interaction strengths from the null search results from XENON1T, PICO-60 and IceCube. For some interactions, the limits on the coupling strengths are relaxed by more than  one order of magnitude. We also present a method that allows to combine the results from different experiments, thus exploiting the synergy between different targets in exploring the parameter space of dark matter-nucleon interactions.}

\begin{document}

\maketitle

\section{Introduction}
\label{sec:introduction}

Dark matter (DM) comprises around 27$\%$ of the mass density of the Universe~\cite{Planck:2018vyg}. However, except for its gravitational effects on the ordinary matter, no other dark matter interaction has been detected up to now (for reviews, see {\it e.g.} \cite{Jungman:1995df,Bertone:2004pz,Bergstrom:2000pn,Feng:2010gw}). There exists a worldwide experimental effort aiming to detect the dark matter interactions with nuclei, chiefly motivated by a well motivated class of dark matter models where the dark matter reached thermal equilibrium with the plasma of Standard Model particles in the very early stages of our Universe (the so-called Weakly Interacting Massive Particles, or WIMPs). In these experiments, a putative flux of dark matter particles  reaches an underground laboratory. A fraction of the dark matter particles then scatter-off a nucleus in a dedicated detector, thus inducing a detectable signal in the form of scintillation light, ionization charges or heat (for reviews, see {\it e.g.} \cite{MarrodanUndagoitia:2015veg}). Unfortunately, and despite the exquisite sensitivity of current experiments, no conclusive signal has been detected up to now. 

The upper limits on the dark matter-nucleus interaction rate can be used to set constraints  on the microphysics of the dark sector and its portal interactions with the Standard Model particles. It is common in the literature to assume that the dark matter interacts with nuclei only via the so-called spin-independent or the spin-dependent interaction, and that the dark matter couples with equal strength to protons and to neutrons. These assumptions are however very restrictive. In fact, in most models the dark matter couples differently to protons and to neutrons (see {\it e.g.}~\cite{isospin_violation_danny, isospin_violation1, isospin_violation2, isospin_violation_gao1, isospin_violation_gao2, isospin_violation_frandsen, isospin_violation_hamaguchi, isospin_violation_belanger, isospin_violation_drozd, isospin_violation_lozano}). Also, in certain models the leading interaction in the effective theory does not correspond to the canonical spin-independent nor the spin-dependent interactions, and may even contain several interaction terms interfering with one another (akin to the Fermi theory of weak interactions, where the vector interaction interferes with the axial interaction). For all these models, the comparison of the published limits with the expectations from models is far from straightforward.  

In~\cite{Anja_2020}, it was developed an analytical method to derive upper limits on the coupling strengths of the non-relativistic effective field theory (NREFT) of dark matter-nucleon interactions, including the effect of operator interference, and which are therefore applicable to all models. The method can also be applied for scenarios where only a subset of the operators arise, as occurs in concrete models. Clearly, in that case the limits become more and more stringent, and reduce to the published limits when only one single interaction is considered. In that paper, the method was applied to derive limits on the coupling strengths from the upper limits on the dark matter capture rate in the solar interior, resulting from the non-observation of a neutrino excess in the IceCube data collected in the direction of the Sun. In this work we apply and generalize this approach to derive limits on the coupling strengths from the non-observation of a significant excess in the number of nuclear recoil events in direct detection experiments, concretely XENON1T~\cite{xenon_1t} and PICO-60~\cite{pico60_2015, pico60_2019}. We also extend the method to combine the results of more than one experiment, thereby exploiting the complementarity of different targets in probing the parameter space of the effective field theory. 

The paper is organized as follows. In Section \ref{sec:scattering_rate} we review the effective field theory approach to dark matter-nucleus interactions, and the calculation of the number of signal events in a direct detection experiment or at a neutrino telescope from the recoil rate and the capture rate in the Sun, respectively. In Section \ref{sec:single} we present our formalism to calculate model independent upper-limits on the coupling strengths of the effective field theory from the non-observation of a signal at one given experiment, and in Section \ref{sec:combined} we generalize this method to combine the results from several experiments. We present our conclusions in Section \ref{sec:conclusion}.

\section{Signal rates in the DM-nucleus non-relativistic Effective Field Theory}
\label{sec:scattering_rate}

Two main strategies have been proposed to probe the dark matter-nucleus interactions: the search for nuclear recoils in a dedicated detector induced by scatterings off dark matter particles in the Solar System~\cite{GoodmanWitten,Jungman:1995df}, and the search for a high energy neutrino flux in the direction of the Sun generated by annihilations of dark matter particles previously captured in the Sun by scatterings with the solar matter~\cite{Silk:1985ax,Srednicki:1986vj,Griest:1986yu}.

The differential rate of nuclear recoils with energy $E_R$  at a direct detection experiment consisting of $N_T$ targets of the nuclear species $T$, due to their interaction with dark matter particles with mass $m_\chi$, is given by (see {\it e.g.} \cite{Lewin:1995rx})
\be
\frac{d R_{\chi T}}{d E_R}= N_T\frac{\rho_{\chi}}{m_{\chi}}\int_{v_{ \rm min}}d^3 v_T f(\vec{v_T}) v_T \frac{d\sigma_T}{d E_R}.
\label{eq:dr_de}
\ee
Here, $v_T\equiv |\vec{v_T}|$ is dark matter speed in the reference frame of the nuclear center of mass, while $v_{\rm min}= \sqrt{\frac{m_T E_R}{2 \mu_{T}^2}}$ is the minimum dark matter speed producing the recoil energy $E_R$, with $\mu_T$ the dark matter-target nucleus reduced mass and $m_T$ the target mass. Further, $\rho_\chi$ and $f(\vec v_T)$ are the local dark matter density and velocity distribution, for which we adopt the values of the Standard Halo Model, namely a local density $\rho_\chi=0.3$  GeV/cm$^3$~\cite{Local_density1} and a velocity distribution with a Maxwell-Boltzmann form,  with velocity dispersion 220 km/s~~\cite{v0_koposov} and truncated at the escape velocity from the Milky Way, 550 km/s~\cite{vesc_2014}. Finally, $d\sigma_T/d E_R$ is the differential cross-section for the dark matter scattering off the target nucleus $T$.

The capture rate, on the other hand, reads \cite{Gould:1987ir}
\begin{align}
C=& \sum_T \int_0^{R_\odot} \, d r \, 4\pi\,r^2\,\eta_T(r)\,\frac{\rho_\chi}{m_\chi}\,\int_{v \leq v_{\text{max},T}^{\text{(Sun)}}(r)} d^3 v \, \frac{ f (\vec{v})}{v}\,w^2(r) \nonumber\\&			 		~\times\int_{m_\chi v^2 /2}^{2 \mu_T^2 w^2(r)/m_{T}} dE_R \, \frac{d \sigma_T}{dE_R}(w(r), E_R) \;.
\end{align}
Here, $\eta_T (r)$ is the number density of the target nucleon $T$ at distance $r$ from the center of the Sun, for which we adopt the Standard Solar Model AGSS09ph \cite{serenelli2009new}. Further, $v$ denotes the dark matter velocity asymptotically far away from the Sun, and $w^2 (r)\,=\,v^2\,+\,v_{\mathrm{esc}}^2 (r)$
is the dark matter velocity at the distance $r$, where $v_{\mathrm{esc}}(r)$ is the escape velocity at that distance.

Since the momentum transfer in the dark matter-nucleus scattering is small compared to the target mass, the scattering can be conveniently described using a NREFT. In this formulation, the interaction Hamiltonian between the target nucleus $T$, and a dark matter particle with spin up to 1/2 reads~\cite{haxton1,haxton2}
\begin{align}\label{eq:H}
	{\cal H}_{\chi N} = \sum_a
	\sum_{i} {\cal C}_i^a \mathcal{O}_i^a\,,
\end{align}
where the index $a$ labels the nucleons in the target nucleus, $a=1, ..., A$, with $A$ the mass number, and $i$ labels the possible Galilean
invariant operators, which depend on the momentum transfer, $\vec{q}$, the DM and nuclear spins, $\vec{S}_{\chi}$ and $\vec{S}_{N}$, and the relative transverse velocity $\vec{v}^\perp=\vec{v}+\vec{q}/2\mu_{N}$. Here, $v$ and $\mu_N$ are dark matter-nucleon relative velocity and reduced mass. As shown in ~\cite{haxton1,haxton2}, there are 14 operators which depend at most linearly on the relative transverse velocity, and which are listed in Table \ref{tab:operators}.
Further, ${\cal C}_i^a$ denotes the coupling strength of the operator $\mathcal{O}_i^a$. Since the nucleon is an isospin doublet, ${\cal C}_i^a$ can be expressed as a $2\times 2$ matrix.  It is common to express
\begin{align}
	{\cal C}_i^a=c_i^0 \mathds{1}^a_{2\times 2}+c_i^1 \mathds{\tau}_3^a\,,
\end{align}
where $\mathds{1}_{2\times 2}^a$ ($\mathds{\tau}_{3}^a$) is the identity (third Pauli matrix) in the $a$-th nucleon isospin space, and $c_i^0$ ($c_i^1$) is the associated isoscalar (isovector) coupling constant. Alternatively, one can cast
\begin{align}
	{\cal C}_i^a=c_i^p (\mathds{1}^a_{2\times 2}+\mathds{\tau}_3^a)
	+c_i^n (\mathds{1}^a_{2\times 2}-\mathds{\tau}_3^a)\;,
\end{align}
where
\begin{align}
	c_i^n&=\frac{1}{2}(c_i^0-c_i^1)\;,\nonumber \\
	c_i^p&=\frac{1}{2}(c_i^0+c_i^1)\;,
\label{eq:np_from_10}
\end{align}
are respectively the coupling constants to the neutron and the proton. From the interaction Hamiltonian, we calculate $d\sigma_T/dE_R$ using the methods described in~\cite{haxton1,haxton2}. We use one-body density matrix elements (OBDMEs) computed in~\cite{haxton2} and implemented in the {\sffamily Mathematica} package {\sffamily DMFormFactor}.  Finally, we use the WimPyDD code~\cite{wimpydd} to calculate the total recoil rate integrating over all possible recoil energies, properly taking into account the detector response, and summing over all possible target species in the detector.

\begin{table}[t!]
\begin{center}
\begin{tabular}{|l|l|}
\hhline{|-|-|}
$ \CO_1 = 1_\chi 1_N$ & $\CO_9 = i \vec{S}_\chi \cdot (\vec{S}_N \times \frac{\vec{q}}{m_N})$ \\
$\CO_3 = i \vec{S}_N \cdot (\frac{\vec{q}}{m_N} \times \vec{v}^\perp)$ & $\CO_{10} = i \vec{S}_N \cdot \frac{\vec{q}}{m_N}$ \\
$\CO_4 = \vec{S}_\chi \cdot \vec{S}_N$ & $\CO_{11} = i \vec{S}_\chi \cdot \frac{\vec{q}}{m_N}$\\
$\CO_5 = i \vec{S}_\chi \cdot (\frac{\vec{q}}{m_N} \times \vec{v}^\perp)$ & $\CO_{12} = \vec{S}_\chi \cdot (\vec{S}_N \times \vec{v}^\perp)$ \\
$\CO_6= (\vec{S}_\chi \cdot \frac{\vec{q}}{m_N}) (\vec{S}_N \cdot \frac{\vec{q}}{m_N})$ & $\CO_{13} =i (\vec{S}_\chi \cdot \vec{v}^\perp  ) (  \vec{S}_N \cdot \frac{\vec{q}}{m_N})$ \\
$\CO_7 = \vec{S}_N \cdot \vec{v}^\perp$ & $\CO_{14} = i ( \vec{S}_\chi \cdot \frac{\vec{q}}{m_N})(  \vec{S}_N \cdot \vec{v}^\perp )$\\
$\CO_8 = \vec{S}_\chi \cdot \vec{v}^\perp$ & $\CO_{15} = - ( \vec{S}_\chi \cdot \frac{\vec{q}}{m_N}) \big((\vec{S}_N \times \vec{v}^\perp) \cdot \frac{\vec{q}}{m_N}\big)$ 
\\ \hline
\end{tabular}
\caption{Non-relativistic Galilean invariant operators for dark matter with spin $1/2$.}
\label{tab:operators}
\end{center}
\end{table}

For the ${\cal O}_1$ and ${\cal O}_4$ operators, corresponding to the spin-independent (SI) and spin-dependent (SD) interactions respectively, it is common to express the coupling strengths $c_i^{\mathscr N}$ with the nucleon ${\mathscr N}$, ${\mathscr N}=n,p$, using instead the cross-sections, 
\begin{align}
 \sigma^{\rm SI}_{\chi{\mathscr N}}&=\frac{(c_1^{\mathscr N})^2\mu_{\chi\mathscr N}^2}{\pi}\;,
 \nonumber\\
 \sigma^{\rm SD}_{\chi\mathscr N}&=\frac{3}{16}\frac{(c_4^{\mathscr N})^2\mu_{\chi\mathscr N}^2}{\pi}\;,
 \label{eq:sdcs}
\end{align}
where  $\mu_{\chi{\mathscr N}}$ is the reduced mass of dark matter-nucleon system. 

For a spin-1/2 dark matter and assuming a contact interaction, there are 28 couplings constants corresponding to $c^p_i$ and $c^n_i$ for $i=1, 3, ..., 15$ in a single vector $\bf c$ with components $c_\alpha$, $\alpha = 1, ..., 28$, which encodes the type and strength of the couplings arising from a concrete dark matter model, upon matching to the effective theory of dark matter-nucleon interactions. Since the Hamiltonian is linear in the coupling strengths, the total number of signal events expected at a given dark matter search experiment ${\mathscr E}$ can be cast as~\cite{Catena_dama,Catena_halo_indep_eft}
\begin{align}
N^{\rm sig}_{\mathscr E}({\bf c})= {\bf c}^T \mathbb{N}_{\mathscr E}{\bf c}\;,
\label{eq:Nsignal}
\end{align}
where $\mathbb{N}_{\mathscr E}$ is a $28\times 28$ real symmetric matrix that depends on the dark matter mass, as well as on the local dark matter density and velocity distribution, and which encodes all the details of the set-up of the experiment ${\mathscr E}$. For a direct detection experiment, these include the nuclear response functions and the exposure, while for a neutrino telescope,  the dark matter annihilation channel and annihilation rate, the solar composition, or the neutrino propagation and flavor conversion inside the Sun. 

\section{Limits on the coupling strengths from a single experiment}\label{sec:single}

Let us consider an experiment ${\mathscr E}$, for which the number of background events is $N_{\mathscr E}^{\rm bck}$, the number of observed events is $N_{\mathscr E}^{\rm obs}$ and the number of expected signal events is  $N_{\mathscr E}^{\rm sig}(\bf c)$, given in Eq.~(\ref{eq:Nsignal}). The $\chi^2$ distribution is given in terms of the experimental likelihood as~\cite{Likelihood_2017}
 \be\label{eq:chi21}
 \chi_{\mathscr E}^2({\bf c})=-2\, {\rm ln}\mathcal{L}\big(N_{\mathscr E}^{\rm sig}({\bf c})\big),
 \ee
 which is in general a complicated function of $N_{\mathscr E}^{\rm sig}(\bf c)$, $N_{\mathscr E}^{\rm bck}$ and $N_{\mathscr E}^{\rm obs}$. On the other hand, for the experiments we will discuss in this paper the $\chi^2$ can be well approximated by a quadratic function
 \begin{align}
 \chi_{\mathscr E}^2({\bf c}) \simeq a_{\mathscr E} (N_{\mathscr E}^{\rm sig})^2+b_{\mathscr E} N_{\mathscr E}^{\rm sig}+c_{\mathscr E}\;,
 \label{eq:fit_chi}
 \end{align}
 with coefficients $a_{\mathscr E}$, $b_{\mathscr E}$, and $c_{\mathscr E}$ listed in Table \ref{tab:exp} (for details, see Appendix \ref{app:likelihoods}).
 
 \begin{table}[t]
\begin{center}
\begin{tabular}{|c|c|c|c|}
\hline 
experiment &  $a_{\mathscr E}$ & $b_{\mathscr E}$ & $c_{\mathscr E}$ \\
\hline
XENON1T &  0.06713 & $-1.072$ & 8.707 \\
PICO-60 (1st bin) &  0.29010   & $-1.728$ & 5.440 \\
PICO-60 (2nd bin) &  0 & 2 & 0 \\
IceCube & 0.001046 & 0.01092 & 8.696 \\
DeepCore & 0.002376 & -0.06191 & 8.298 \\
\hline
\end{tabular}
\end{center}
\caption{Parameters of the approximate $\chi^2$  given in Eq.~(\ref{eq:fit_chi}) for the XENON1T, PICO-60, and IceCube/DeepCore experiments.}
\label{tab:exp}
\end{table}

The 90\% C.L. allowed region for the coupling strengths is determined by the condition $\chi^2_{\mathscr E}-\chi^2_{\mathscr E,\rm min}\leq 2.71$,\footnote{Here, we have used that the difference in dimensionality between $\chi^2_{\mathscr E}$ and $\chi^2_{\mathscr E,\rm min}$ is equal to one.  If there are no events observed in an experiment with $\chi^2_{\mathscr E,\rm min}=0$ for every dark matter mass, the difference in dimensionality is instead equal to two, giving $\chi^2_{\mathscr E}-\chi^2_{\mathscr E,\rm min}\leq 4.6$.} where $\chi^2_{\mathscr E,\rm min}$ is obtained by minimizing $\chi^2_{\mathscr E}$ with respect to the number of signal events, giving
\be
 \chi_{\mathscr E,\rm min}^2 = c_{\mathscr E}-\frac{b_{\mathscr E}^2}{4 a_{\mathscr E}}.
 \label{eq:comb_chi2min}
\ee

In the theoretical interpretation of null search results, it is common in the literature to consider a single operator at a time, and to derive using the above procedure a 90\% C.L. upper limit on the corresponding effective coupling. This approach, chiefly motivated to reduce the number of free parameters in the analysis, does not allow a straightforward comparison of the experimental results with the expectations of specific models, which in general predict more than one operator in the effective theory, possibly interfering with one another (a notable example are those models where the dark matter couples to both protons and neutrons, with different strengths). As a result, wrong conclusions about the viability of certain models could be drawn~\cite{Catena_Gondolo_global_limits, Catena_Gondolo_global_fits, Anja_2020}. 

\begin{figure}[t!]
\begin{center}
  \includegraphics[width=0.55\textwidth]{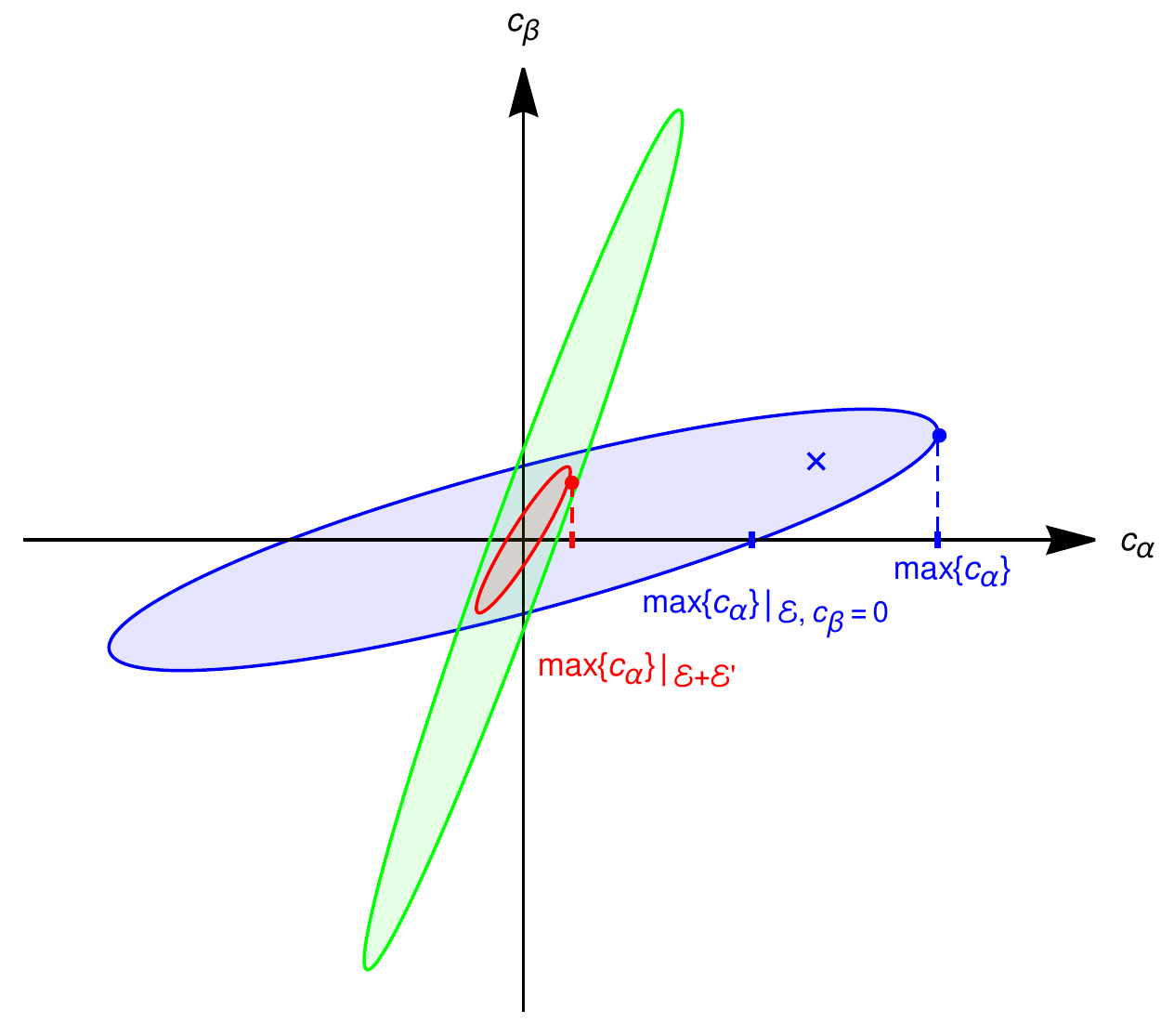}
\end{center}
\caption{Sketch of the 90\% C.L. allowed regions by the experiments $\mathscr{E}$ (blue) and  $\mathscr{E}^\prime$ (green) in the two-dimensional parameter space spanned by coupling strengths $c_\alpha$ and $c_\beta$. The point indicated by the blue cross is allowed by the experiment $\mathscr{E}$, but seemingly ruled out by the requirement $c_{\alpha}<{\rm max}\{c_\alpha\}|_{c_\beta=0}$ (as done implicitly in many analyses). Instead, the requirement $c_{\alpha}<{\rm max}\{c_\alpha\}$ should be imposed. The figure also illustrates the complementarity of experiments in probing the parameter space of the NREFT: the point indicated by the blue cross is allowed by experiment $\mathscr{E}$, but ruled out by experiment $\mathscr{E}^\prime$. Morevoer, the region of the parameter space allowed by the  combined 90\% C.L. limit of both experiments (indicated by the red region) is significantly smaller than the regions allowed by $\mathscr{E}$ and $\mathscr{E}^\prime$ separately, and somewhat smaller than the intersection of the allowed regions by $\mathscr{E}$ and $\mathscr{E}^\prime$.}
\label{fig:allowed_regions_demo}
\end{figure}

This is illustrated in Fig.~\ref{fig:allowed_regions_demo}, which schematically shows the  parameter space allowed by the experiment $\mathscr E$ in the plane spanned by two coupling strengths $c_\alpha$ and $c_\beta$. Assuming that only the coupling $c_\alpha$ is non-zero, the non-observation of a signal at the experiment  ${\mathscr E}$ leads to the {\it model dependent} limit on the coupling strength ${\rm max}\{c_\alpha\}|_{ {\mathscr E},c_\beta=0}$. Let us now consider a  model predicting the coupling strengths indicated by a blue cross. This point, clearly allowed by the experiment ${\mathscr E}$, has $c_\alpha^{\times}>{\rm max}\{c_\alpha\}|_{ {\mathscr E},c_\alpha=0}$ and would therefore be wrongly ruled out. A model independent limit would be given by ${\rm max}\{c_\alpha\}|_{ {\mathscr E}}\equiv
c^{\rm max}_\alpha$, also shown in the Figure, and which excludes any model predicting $c_\alpha>c^{\rm max}_\alpha$  at the 90\% C.L. In particular, this prescription does not exclude the model represented by the blue cross.

To determine $c^{\rm max}_\alpha$ for the experiment ${\mathscr E}$ we pursue and generalize the approach presented in \cite{Anja_2020}. We construct the Lagrangian
\be
	L\,=\,c_\alpha-\lambda \Big [\chi_{\mathscr E}^2({\bf c})-\chi_{{\mathscr E}, \rm min}^2  - 2.71\Big ],
	\label{eq:single_lagrangian}
\ee
where $\lambda$ is a Lagrange multiplier that enforces the requirement that the coupling strengths saturate the 90\% C.L. limit. The coupling strengths ${\bf c}^{\rm max}$ that yield the maximum value for $c_\alpha$ and the Lagrange multiplier $\lambda$ are obtained from extremizing the Lagrangian:
\begin{align}
\frac{\partial L}{\partial c_\beta}\Big|_{{\bf c}={\bf c}^{\rm max}} \,&=\,\delta_{ \beta \alpha}\,-\,2\lambda 
\Big[2 a_{\mathscr E}N_{\mathscr E}^{\rm sig}({\bf c}^{\rm max})+b_{\mathscr E}\Big]
(\mathbb{N}_{\mathscr E})_{\beta \gamma} c_\gamma^{\rm max}  =\,0\;,
\label{eq:comb_max_zeta-equil}\\
\frac{\partial L}{\partial\lambda}\Big|_{{\bf c}={\bf c}^{\rm max}}\,&=
  -\Big[a_{\mathscr E} \Big(N_{\mathscr E}^{\rm sig}({\bf c}^{\rm max})\Big)^2+b_{\mathscr E} N_{\mathscr E}^{\rm sig}({\bf c}^{\rm max})+c_{\mathscr E} -\chi_{\mathscr E,{\rm min}}^2-2.71\Big]=0\;.
\label{eq:comb_max_lambda-equil}
\end{align}

 From the first equation, one obtains an implicit equation for the $\beta$-th coordinate of ${\bf c}^{\rm max}$
\begin{align}
	c_\beta^{\rm max}\, &=\frac{1}{2\lambda \Big[2 a_{\mathscr E}N_{\mathscr E}^{\rm sig}({\bf c}^{\rm max})+b_{\mathscr E}\Big]}(\mathbb{N}^{-1}_{\mathscr E})_{\beta\alpha}.
	\label{eq:cmax(gamma)}
\end{align}
Substituting in Eq.~(\ref{eq:Nsignal}) one obtains the following relation between the maximal number of signal events at the experiment ${\mathscr E}$ and the Lagrange multiplier
\begin{align}
 N_{\mathscr E}^{\rm sig}({\bf c}^{\rm max})=
\frac{1}{4\lambda^2 \Big[2 a_{\mathscr E}N_{\mathscr E}^{\rm sig}({\bf c}^{\rm max})+b_{\mathscr E}\Big]^2} (\mathbb{N}^{-1}_{\mathscr E})_{\alpha\alpha}.
\label{eq:Nsignalmax}
\end{align}
Lastly, from Eq.~(\ref{eq:Nsignalmax}) and Eq.~(\ref{eq:comb_max_lambda-equil}), one obtains $\lambda$ and $N_{\mathscr E}^{\rm sig}({\bf c}^{\rm max})$, which can be substituted in Eq.~(\ref{eq:cmax(gamma)}) to obtain ${\bf c}^{\rm max}$. Specifically, the quantity of interest,  $c_\alpha^{\rm max}$, reads
\begin{align}
c_\alpha^{\rm max}=\sqrt{N^{\rm sig}_{\mathscr E}({\bf c}^{\rm max})(\mathbb{N}_{\mathscr E}^{-1})_{\alpha\alpha}}\;,
\label{eq:cmax_single}
\end{align}
with  $N^{\rm sig}_{\mathscr E}({\bf c}^{\rm max})$ calculable from solving  Eq.~(\ref{eq:comb_max_lambda-equil}). \footnote{The result for a general likelihood function, so that the $\chi^2$ cannot be approximated by a quadratic function of the form Eq.~(\ref{eq:fit_chi}), can be found in Appendix \ref{app:general_formalism}.}

\begin{figure}[t!]
\begin{center}
  \includegraphics[width=0.4\textwidth]{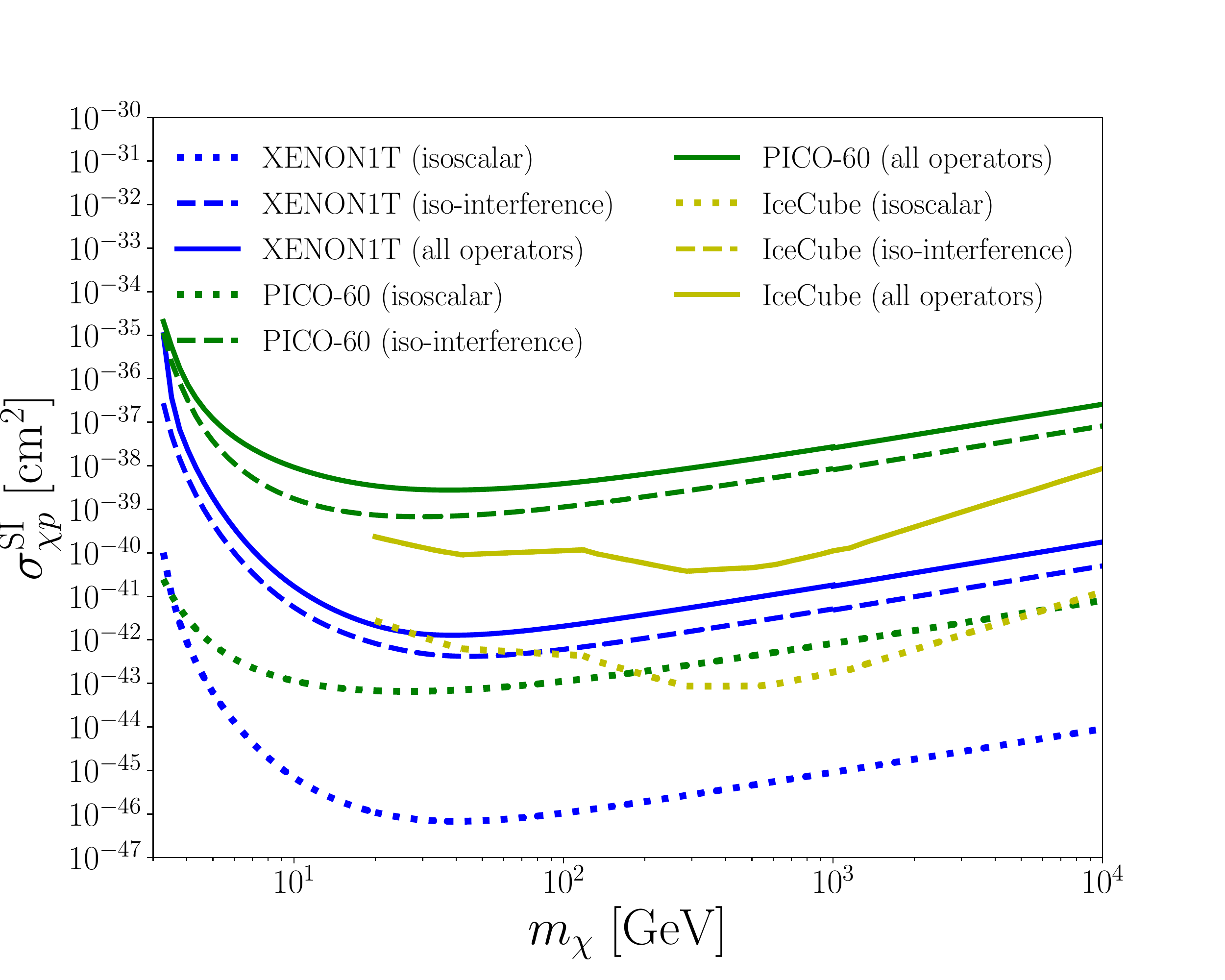}
  \includegraphics[width=0.4\textwidth]{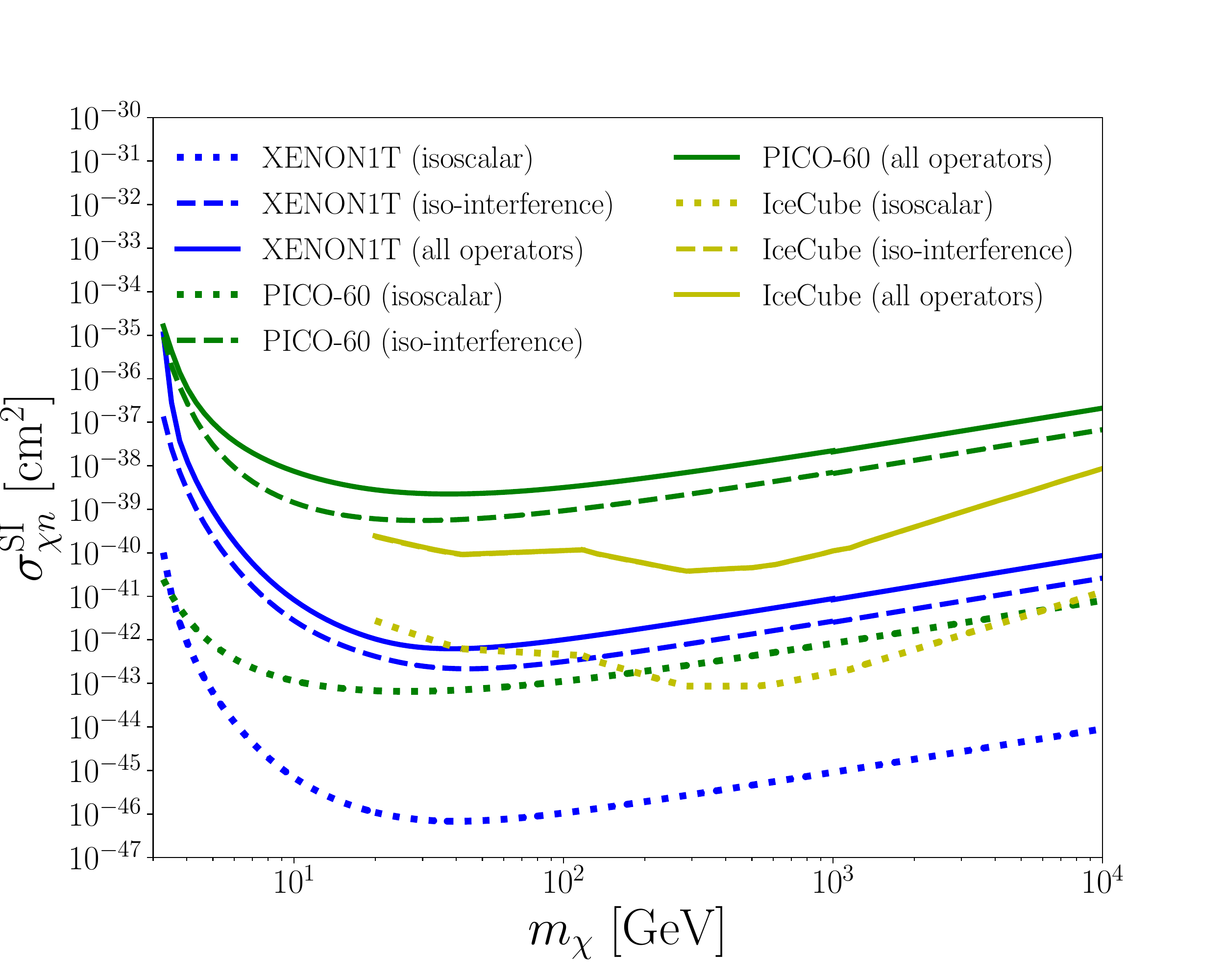}\\
  \includegraphics[width=0.4\textwidth]{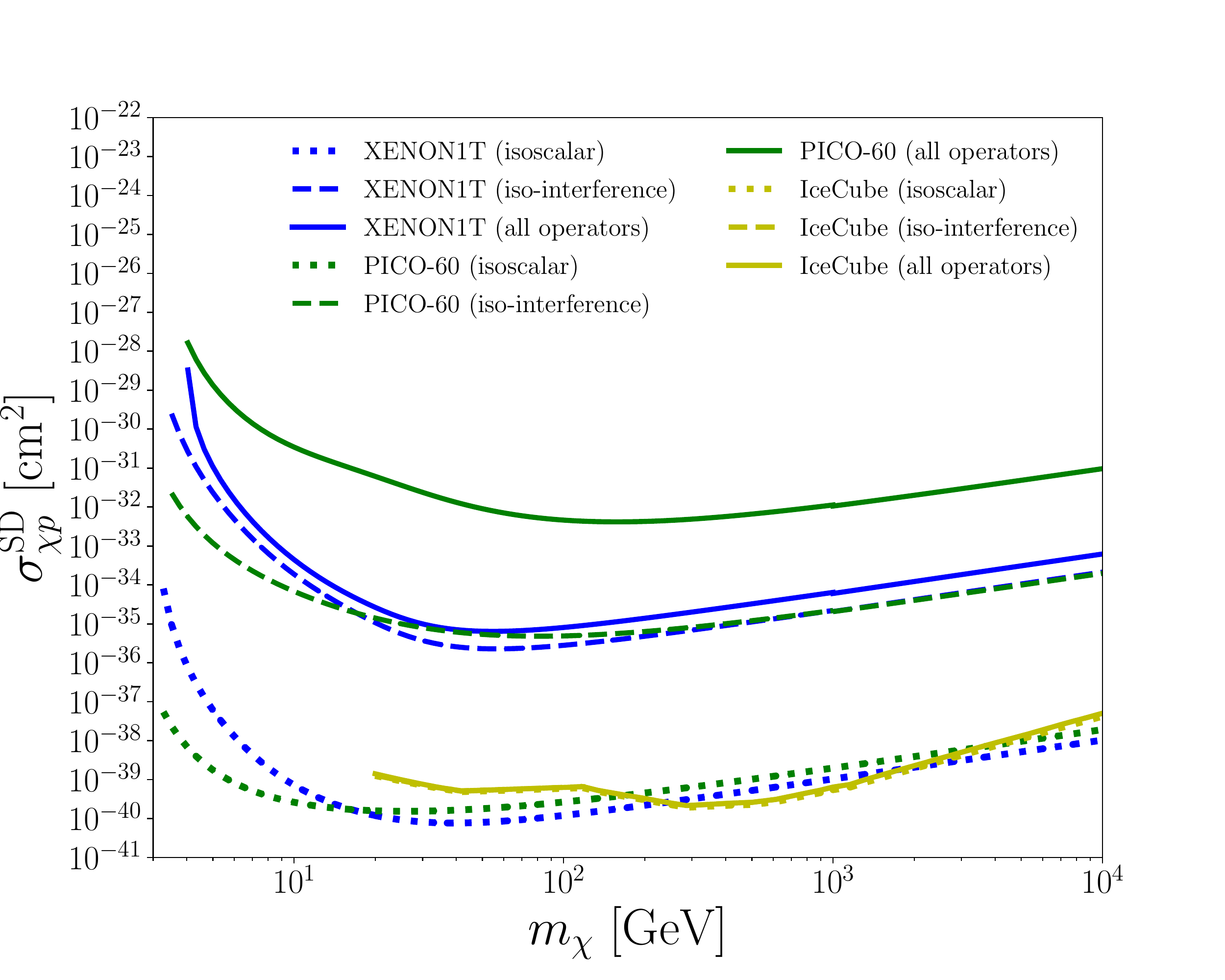}
  \includegraphics[width=0.4\textwidth]{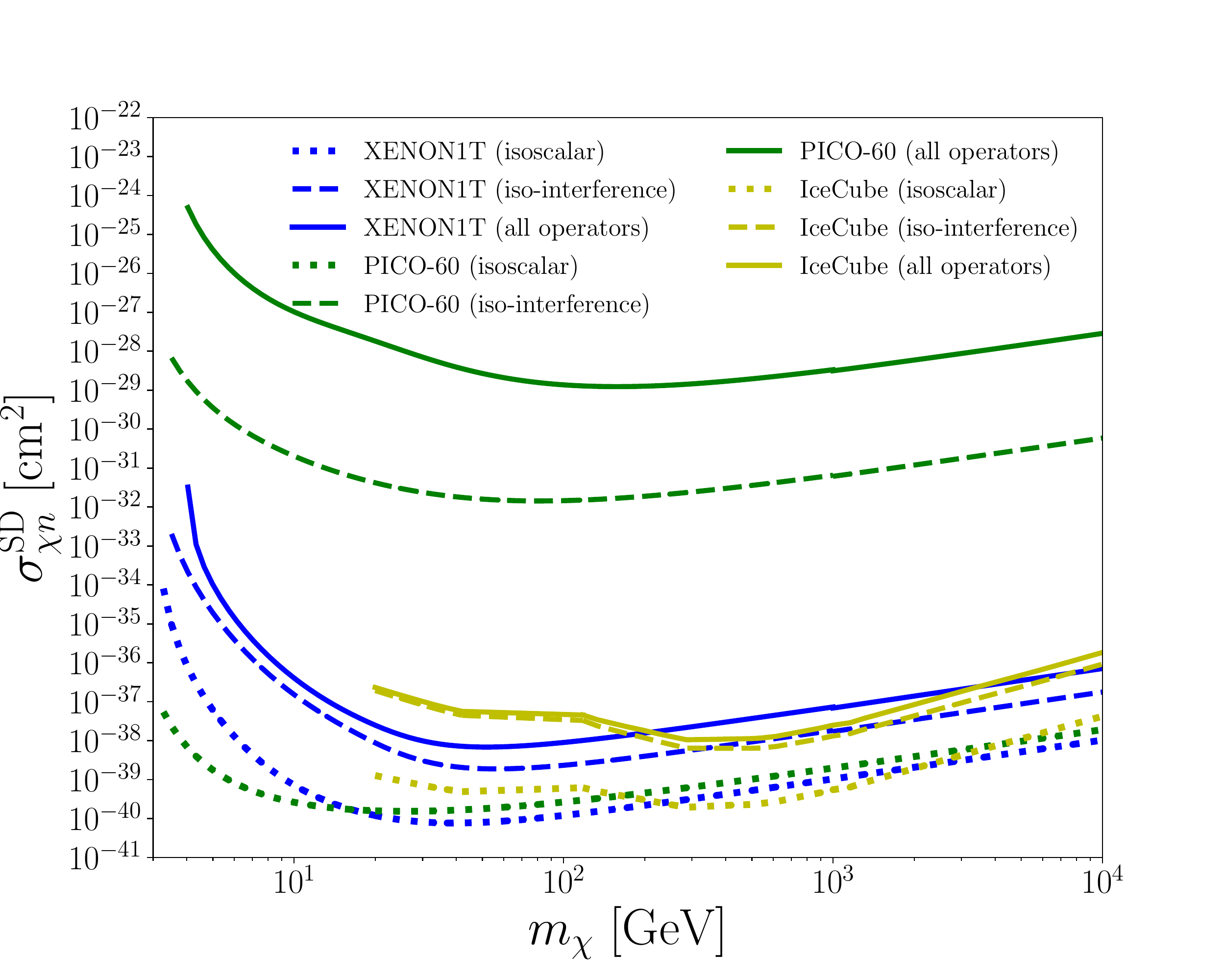}\\
\end{center}
\caption{$90\%$ C.L upper limits on the spin-independent (operator ${\cal O}_1$, top panels) or spin dependent (operator ${\cal O}_4$, bottom panels)  dark matter-proton (left panels) and dark matter-neutron (right panels) cross-section from the non-observation of a signal at XENON1T (blue), PICO-60 (green) and IceCube (yellow) experiments. The dotted lines show the limit considering only the isoscalar interaction in the ${\cal O}_1$ and ${\cal O}_4$ Galilean invariant operators for the SI and SD interactions, respectively; the dashed lines include the effect of interference between the isoscalar and isovector interactions, and the  dotted lines show the limits including all interactions and all Galilean invariant operators.}
  \label{fig:exclusions_SI_SD}
\end{figure}

We show in Fig.~\ref{fig:exclusions_c_single} the model independent upper limits on each of the effective DM-proton (DM-neutron) coupling strengths $c^p_i$ ($c^n_i$) for $i = 1, 3, ..., 15$ as a function of the dark matter mass, from the non-observation of a signal at XENON1T (blue), PICO-60 (green) or IceCube (yellow). For the IceCube experiment, we assume DM annihilations into $W^+W^-$ for $m_{\chi}>100$ GeV and $\tau^+\tau^-$ for $m_\chi<100$ GeV.
We show as a dotted line the limit calculated assuming  that only the isoscalar interaction is present ({\it i.e.} $c^0_j=0$ for $j\neq i$ and $c^1_j=0$ for all $j$), as a dashed line the limit  assuming that the isoscalar and the isovector interactions can interfere for the operator ${\cal O}_i$ ({\it i.e.} $c^0_j=0$,   $c^1_j=0$ for $j\neq i$), and as a solid line the limit assuming that all interactions can be present and interfere with one another ({\it i.e.} no restriction on $c^0_j$ or   $c^1_j=0$ for any $j$). Further, and in order to compare our results with the published limits, we show in Fig. \ref{fig:exclusions_SI_SD} the limits for the coupling strengths of the operators ${\cal O}_1$ and ${\cal O}_4$ recast into spin-independent (top panels) and spin-dependent (bottom panels) DM-proton (left panels) and DM-neutron (right panels) interaction cross-sections, using Eq.~(\ref{eq:sdcs}).
\begin{figure}[t!]
\begin{center}
\includegraphics[trim=0 200 0 00,clip, width=0.90\textwidth]{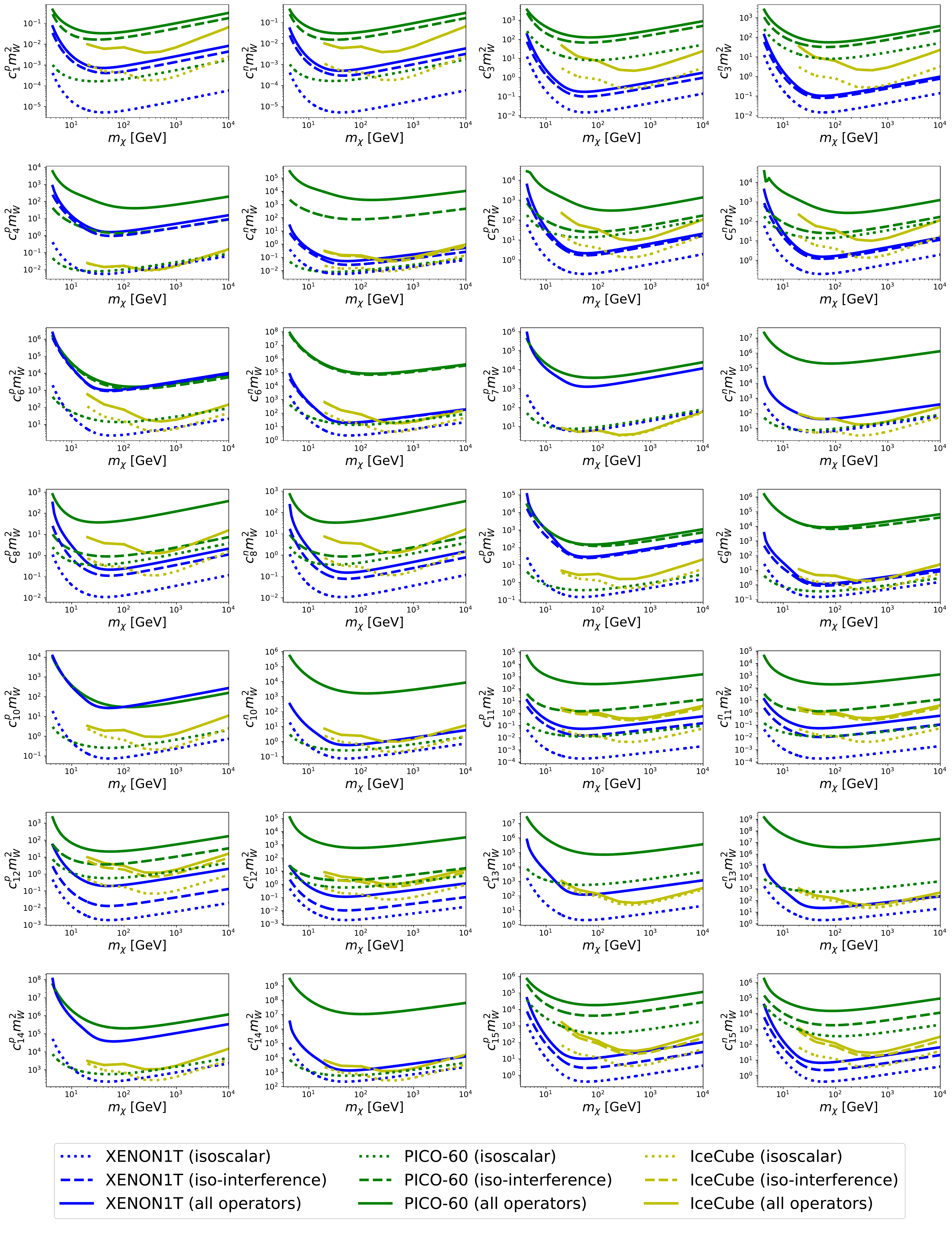}
\end{center}
\caption{90$\%$  C.L upper limits on the coupling strengths of the NREFT of dark matter-nucleon interactions from the null search results from XENON1T (blue), PICO-60 (green) and IceCube (yellow). The dotted line shows the limit assuming only the isoscalar coupling (as published by the experiments), the dashed-line shows the limit resulting from the interference of the isoscalar and isovector interactions for a given Galilean invariant operator ${\cal O}_i$, and the solid line shows the limit resulting from the interference of all interactions and all operators.}\label{fig:exclusions_c_single}
\end{figure}

For most of the interactions, the effect of the interference between the isoscalar and the isovector interaction can be very significant, and the upper limits on the coupling strengths can be substantially relaxed compared to those obtained under the common assumption that the interaction is isoscalar. This conclusion holds in particular for the commonly used SI and SD interactions. For the SI interaction, the limits on the DM-proton and DM-neutron cross-sections from XENON1T and from PICO-60 are relaxed by $\sim$ 4 orders of magnitude for $m_\chi>100$ GeV. For the SD interaction, the limit on the DM-proton cross-section from XENON1T is relaxed by $\sim$ 4 orders of magnitude, and from PICO-60 by $\sim$ 9 orders of magnitude. Also, the limit on the DM-neutron cross-section from XENON1T is relaxed by $\sim$ 2 orders of magnitude, while the limit from PICO-60 is relaxed by $\sim$ 13 orders of magnitude. Clearly, for models where the DM couples to protons and to neutrons with different strength, the predicted values cannot be confronted with the published isoscalar limits. For other experiments, the qualitative conclusion remains, although with quantitative differences. 

\begin{figure}[t!]
	\begin{center}
		\includegraphics[width=0.28\textwidth]{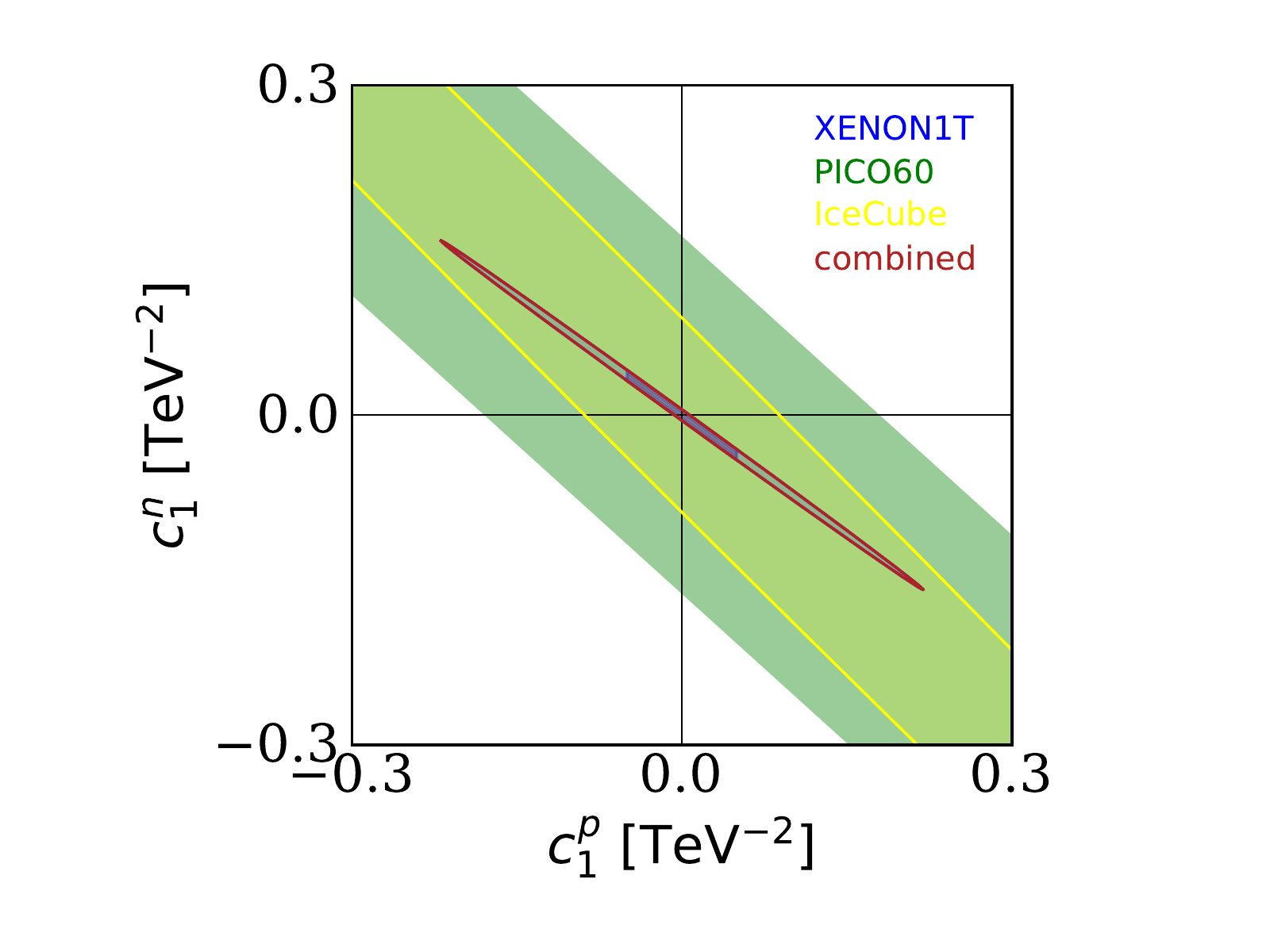}
		\hspace{-0.9cm}
		\includegraphics[width=0.28\textwidth]{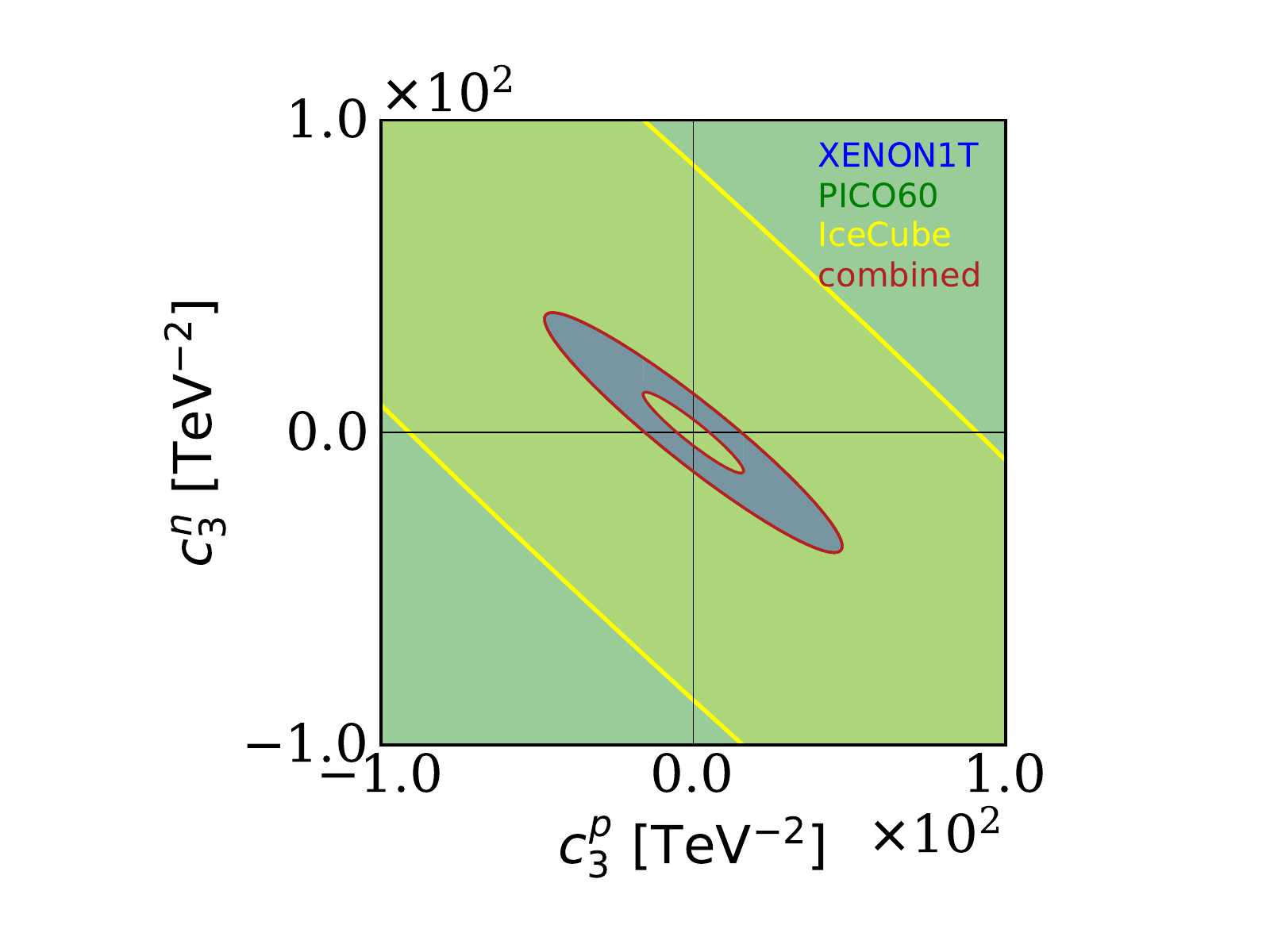}
		\hspace{-1.01cm}
		\includegraphics[width=0.28\textwidth]{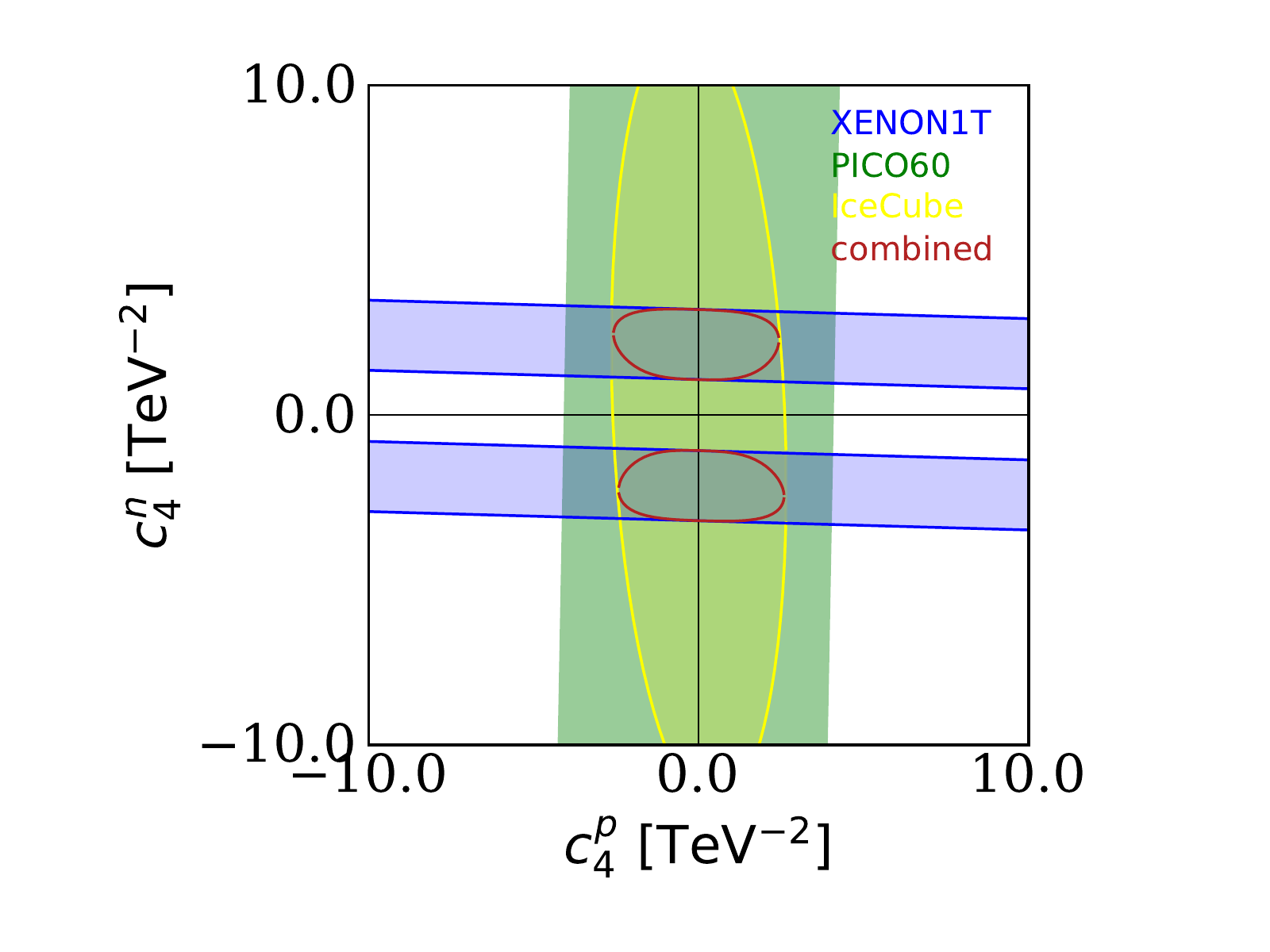}
		\hspace{-1.01cm}
		\includegraphics[width=0.28\textwidth]{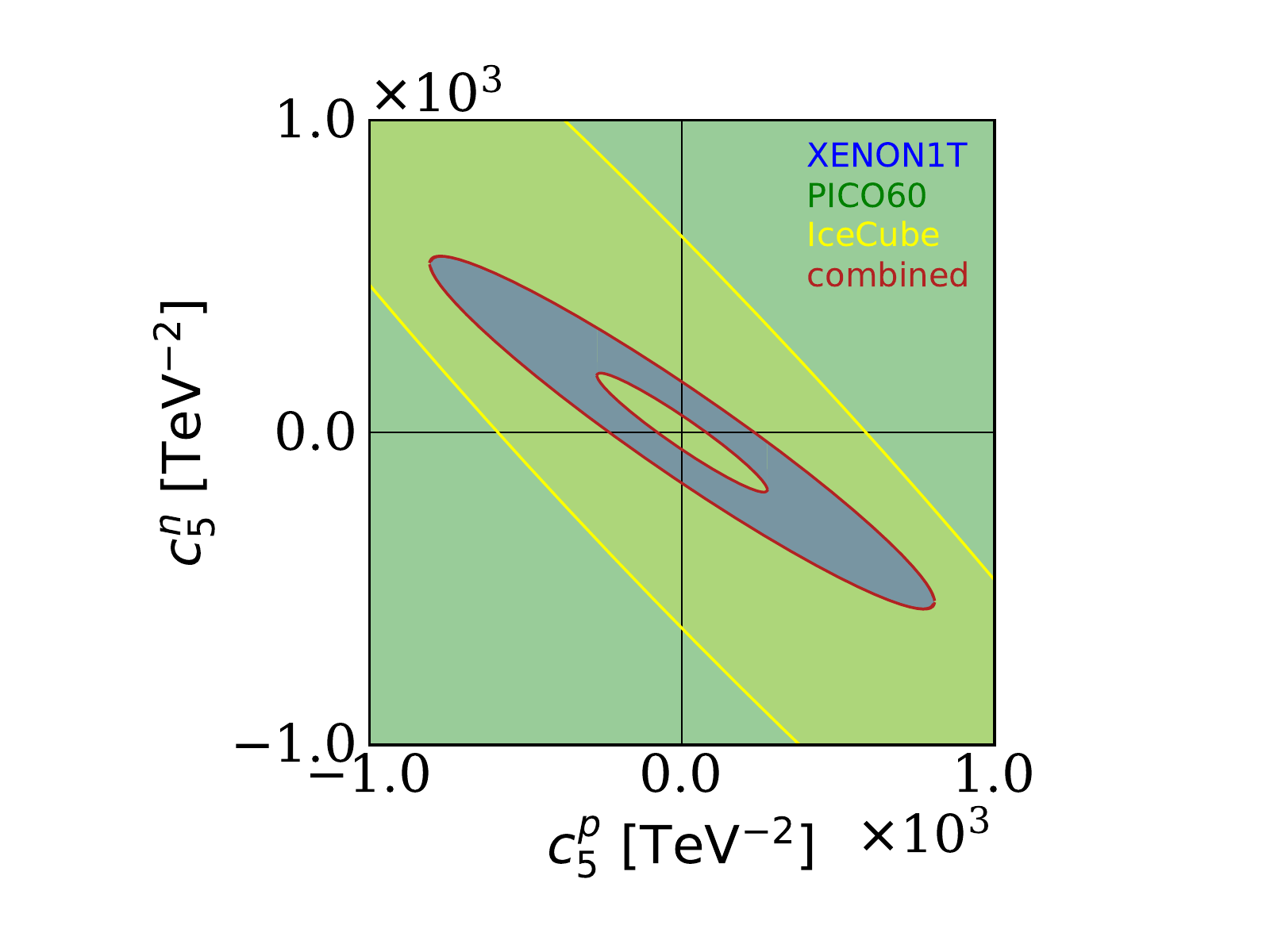}\\
		\includegraphics[width=0.28\textwidth]{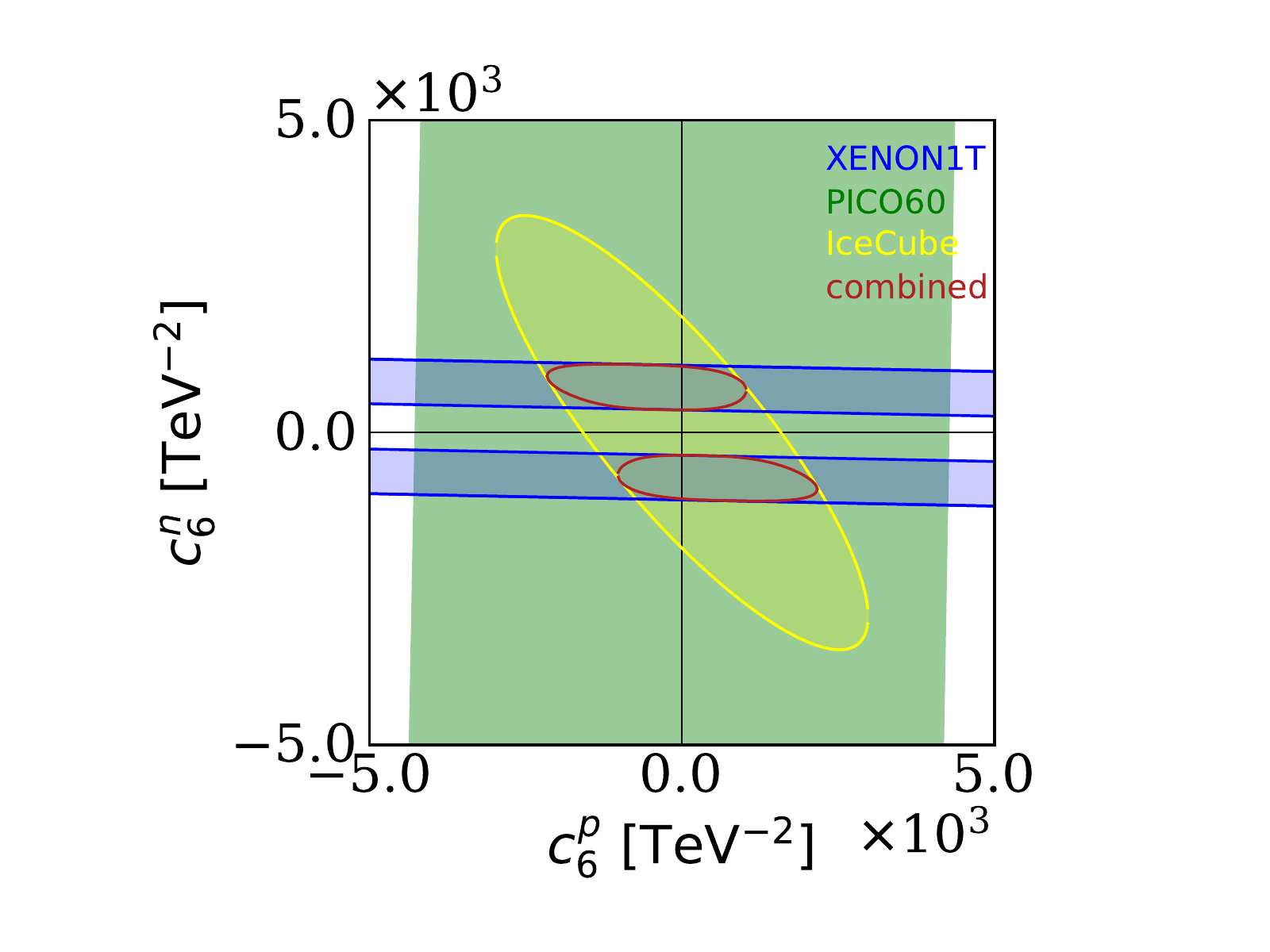}
		\hspace{-0.9cm}
		\includegraphics[width=0.28\textwidth]{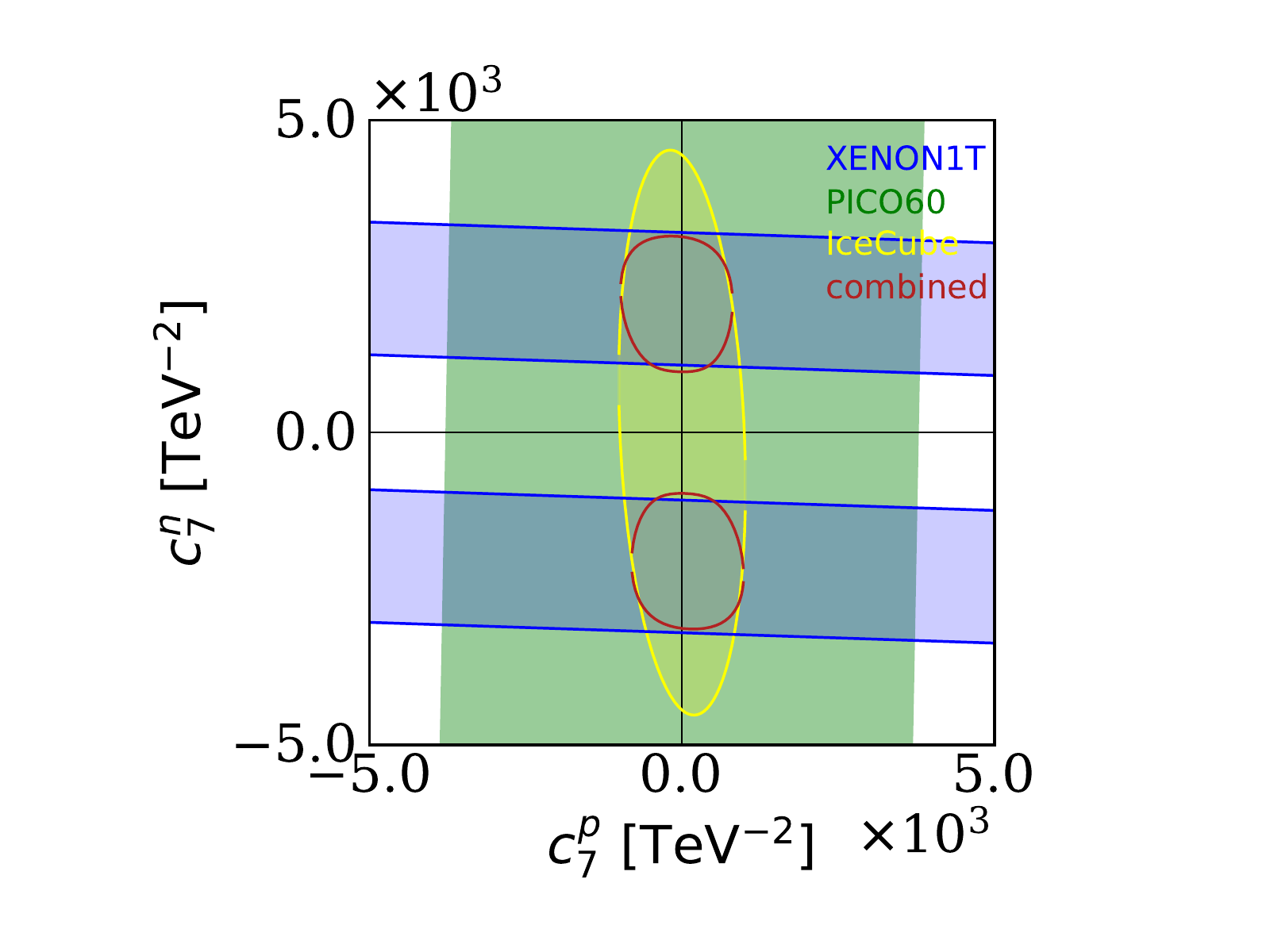}
		\hspace{-0.9cm}
		\includegraphics[width=0.28\textwidth]{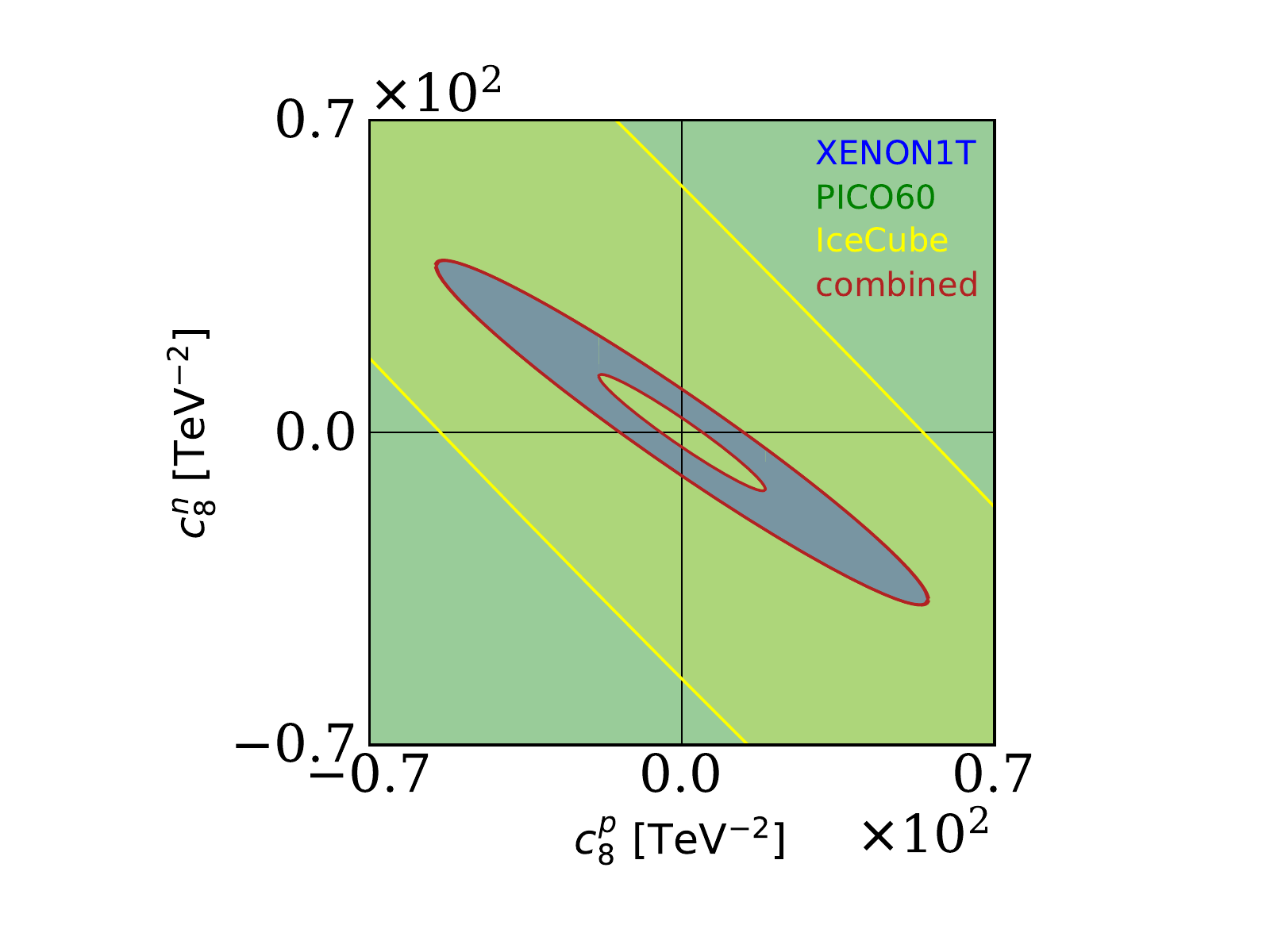}
		\hspace{-1.01cm}
		\includegraphics[width=0.28\textwidth]{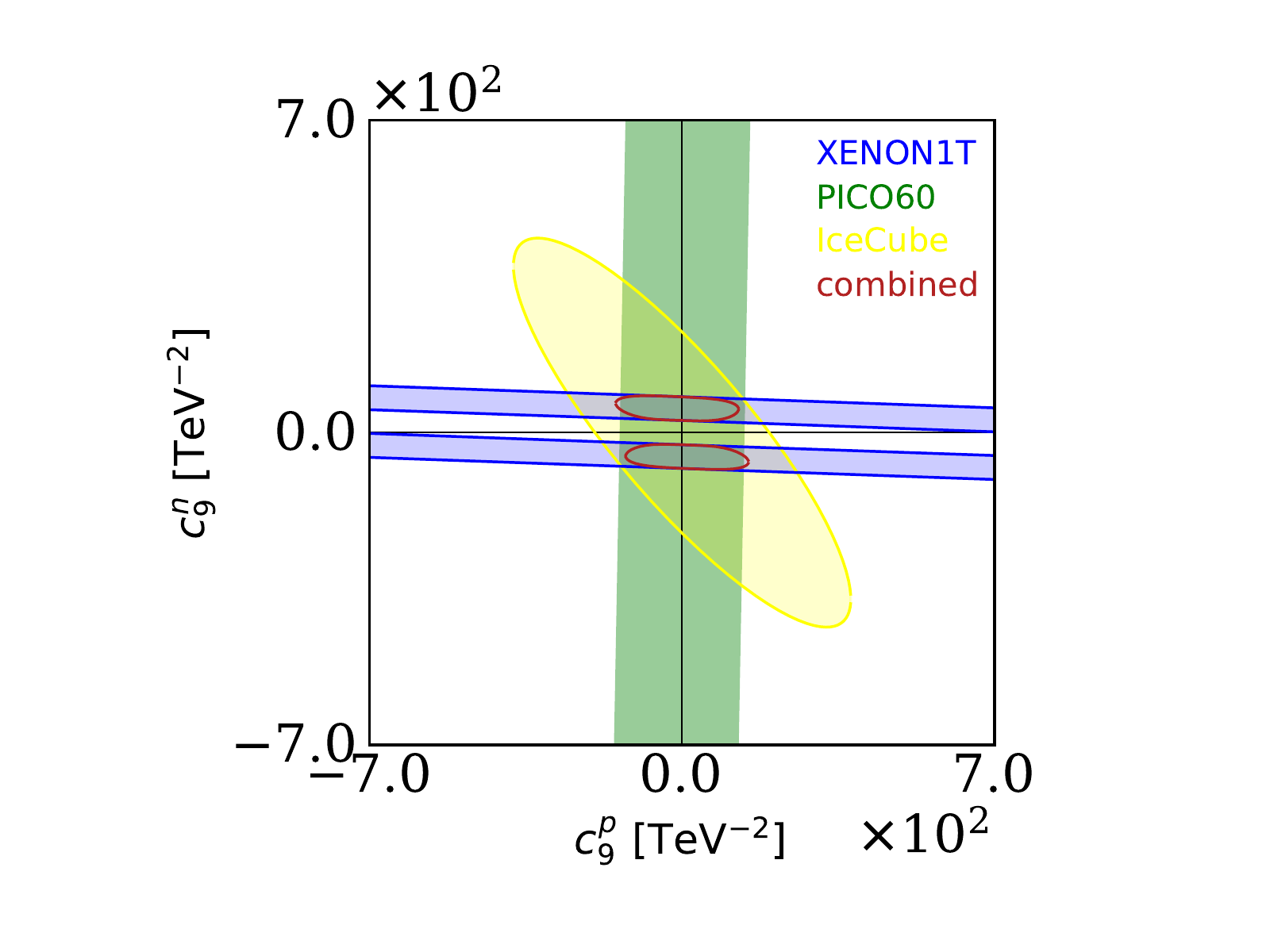}\\
		\includegraphics[width=0.28\textwidth]{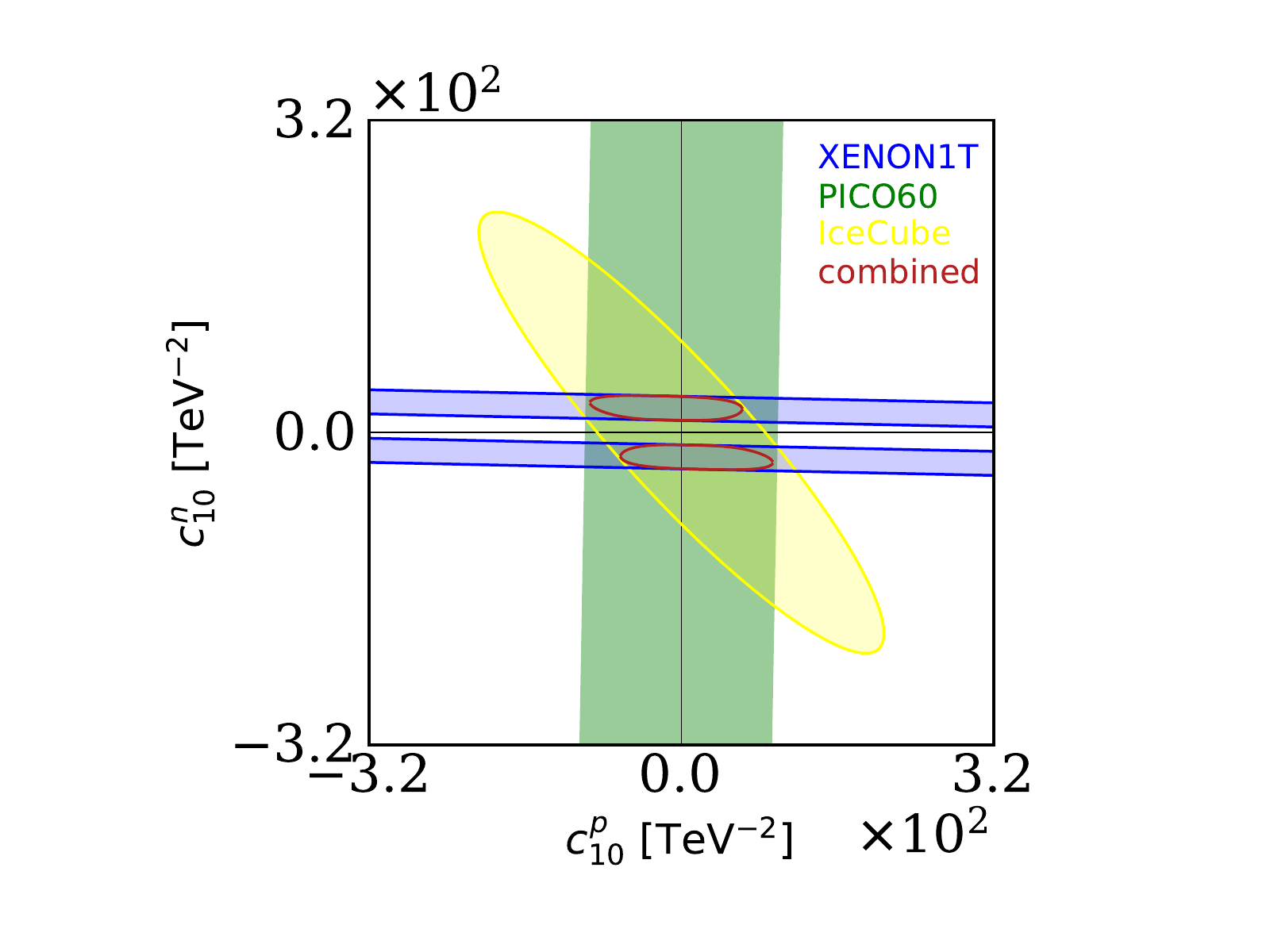}
		\hspace{-0.9cm}
		\includegraphics[width=0.28\textwidth]{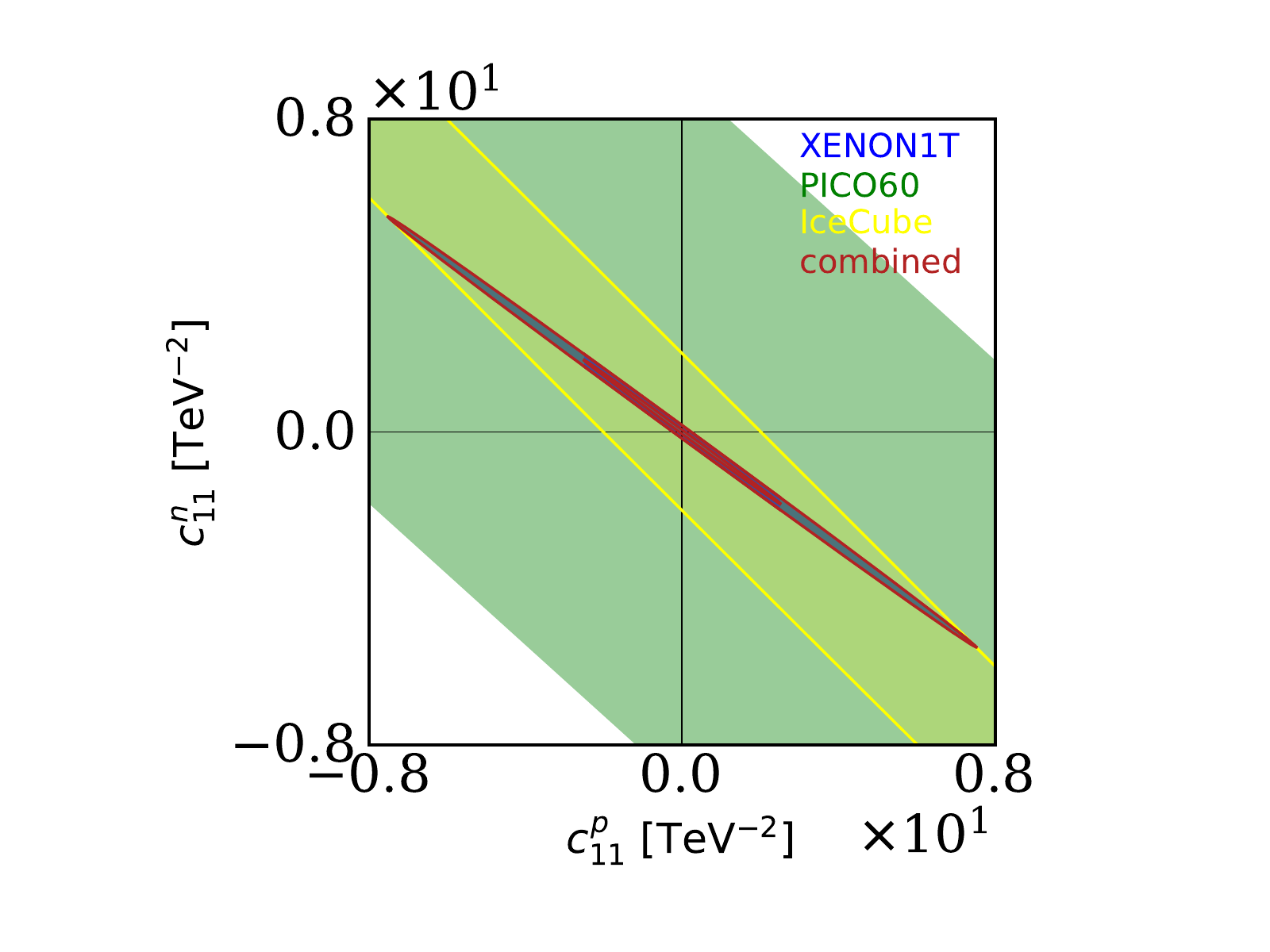}
		\hspace{-1.01cm}
		\includegraphics[width=0.28\textwidth]{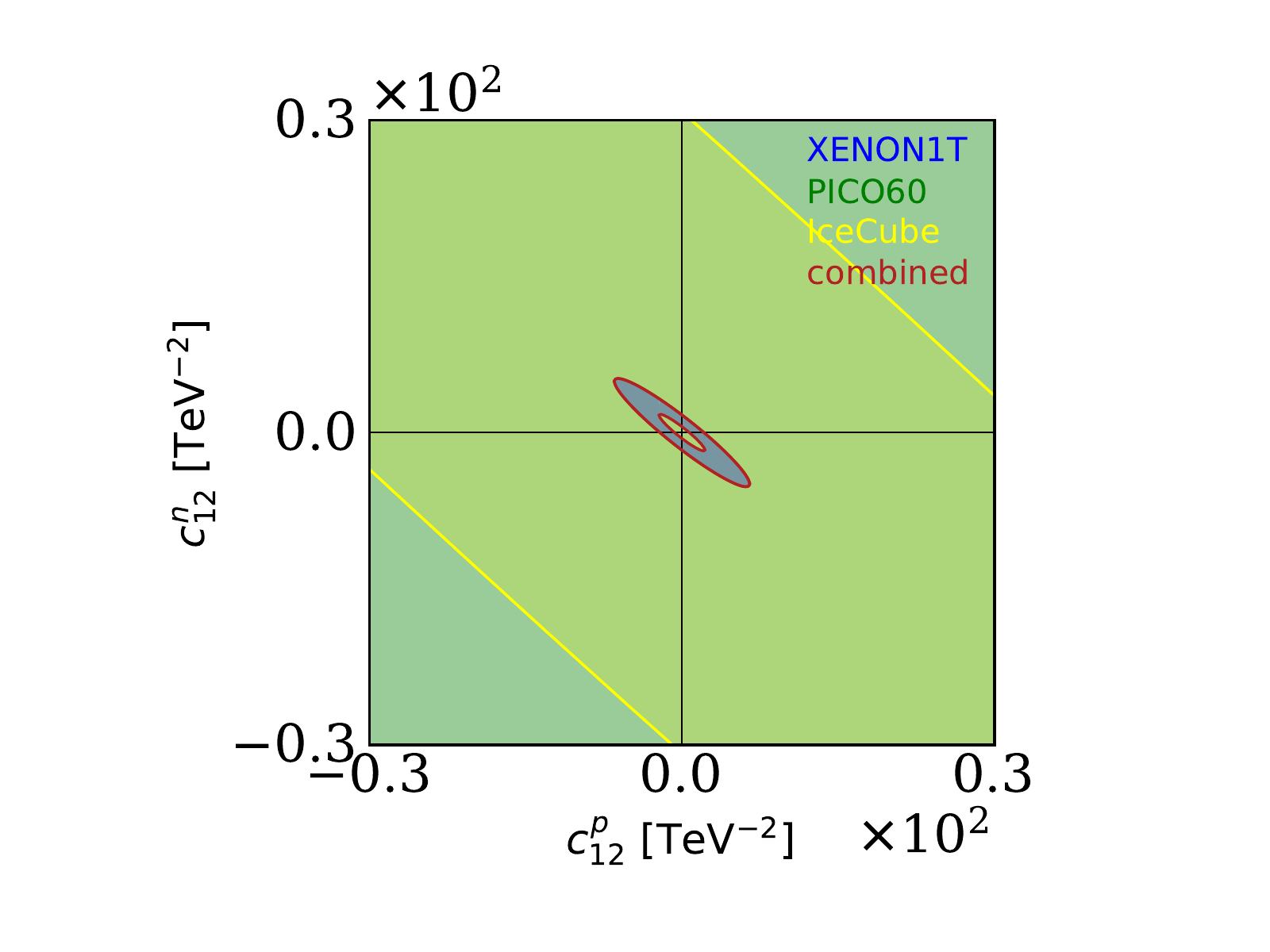}
		\hspace{-1.01cm}
		\includegraphics[width=0.28\textwidth]{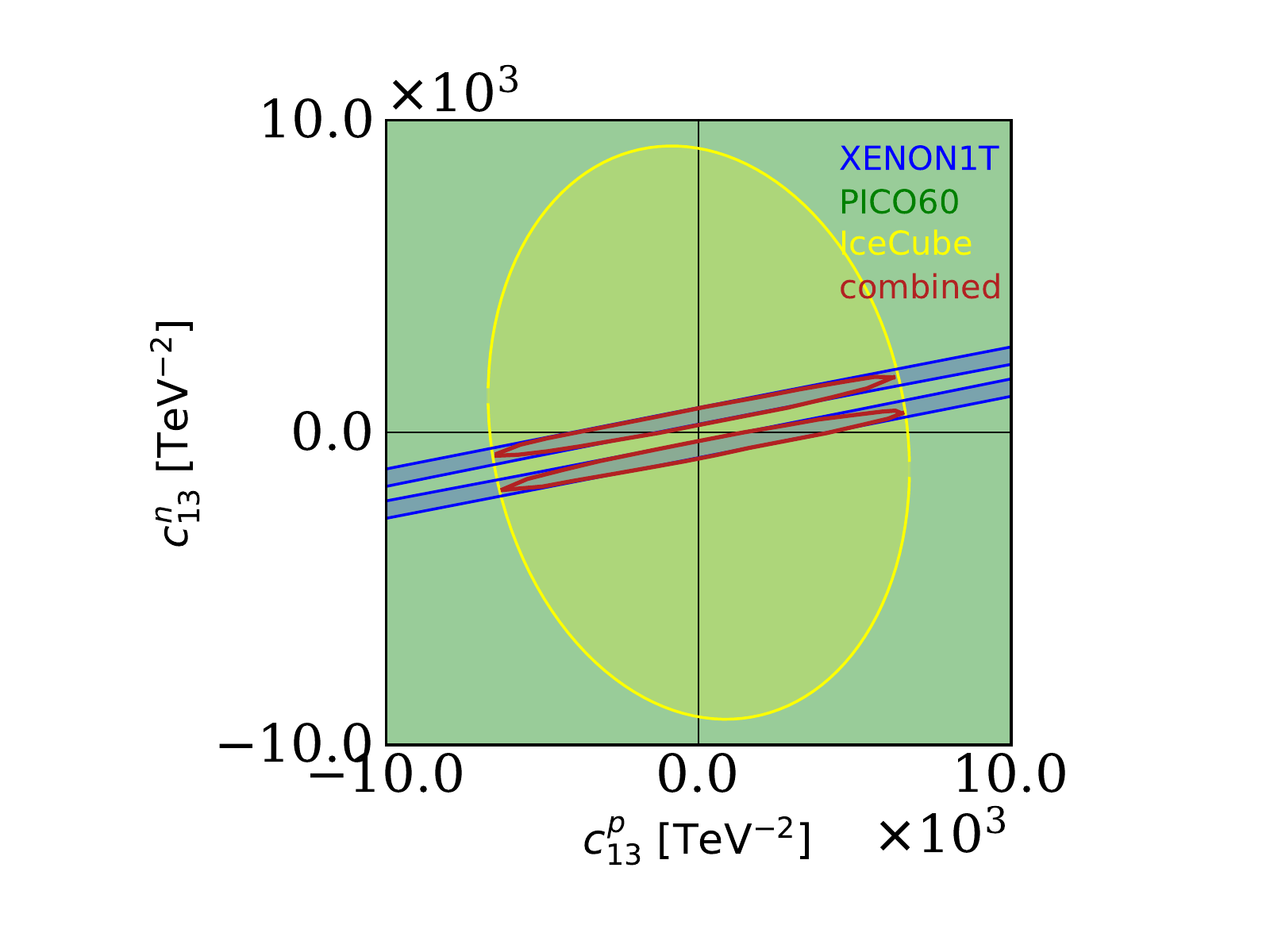}\\
		\includegraphics[width=0.28\textwidth]{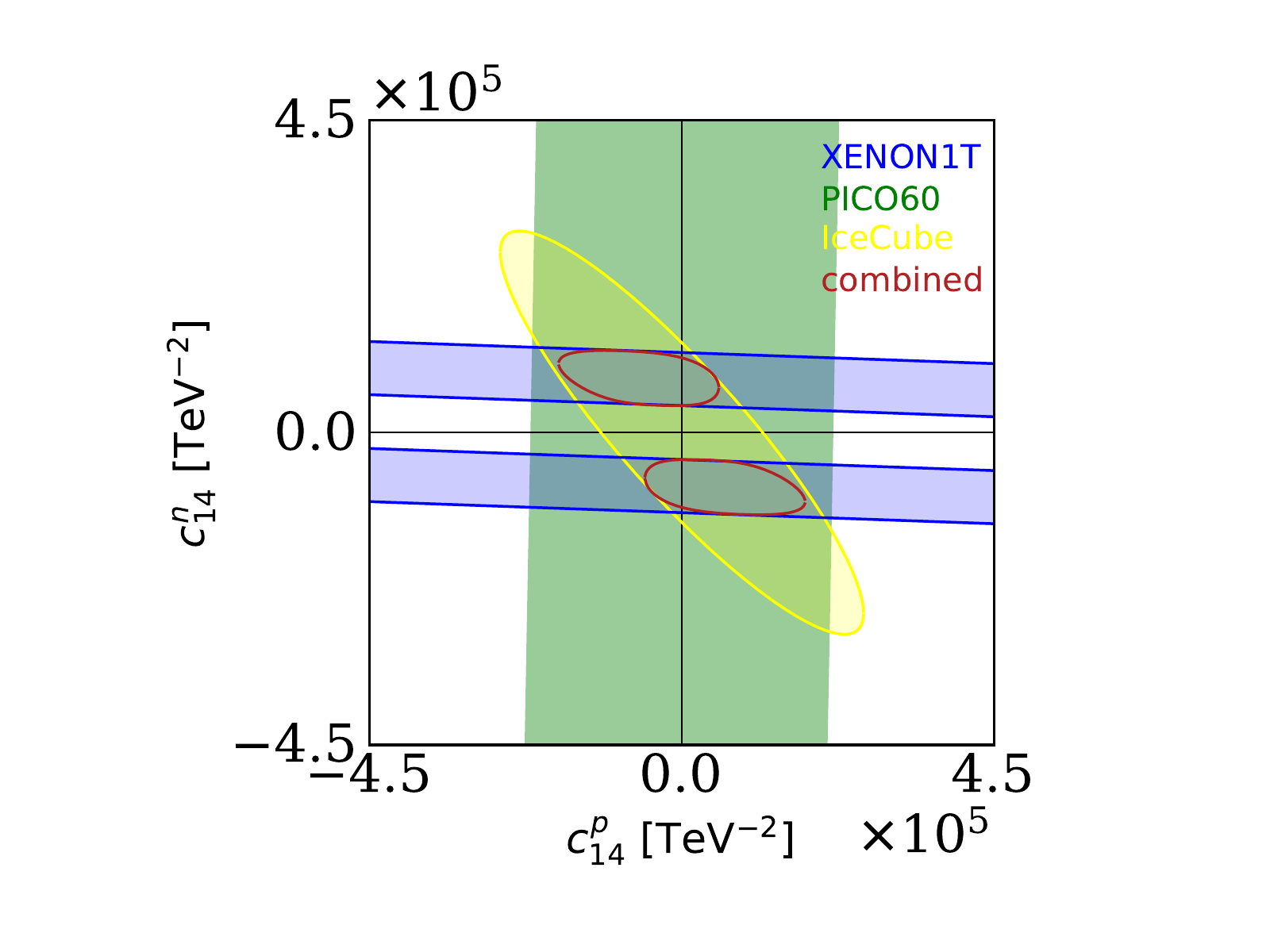}
		\hspace{-0.8cm}
		\includegraphics[width=0.28\textwidth]{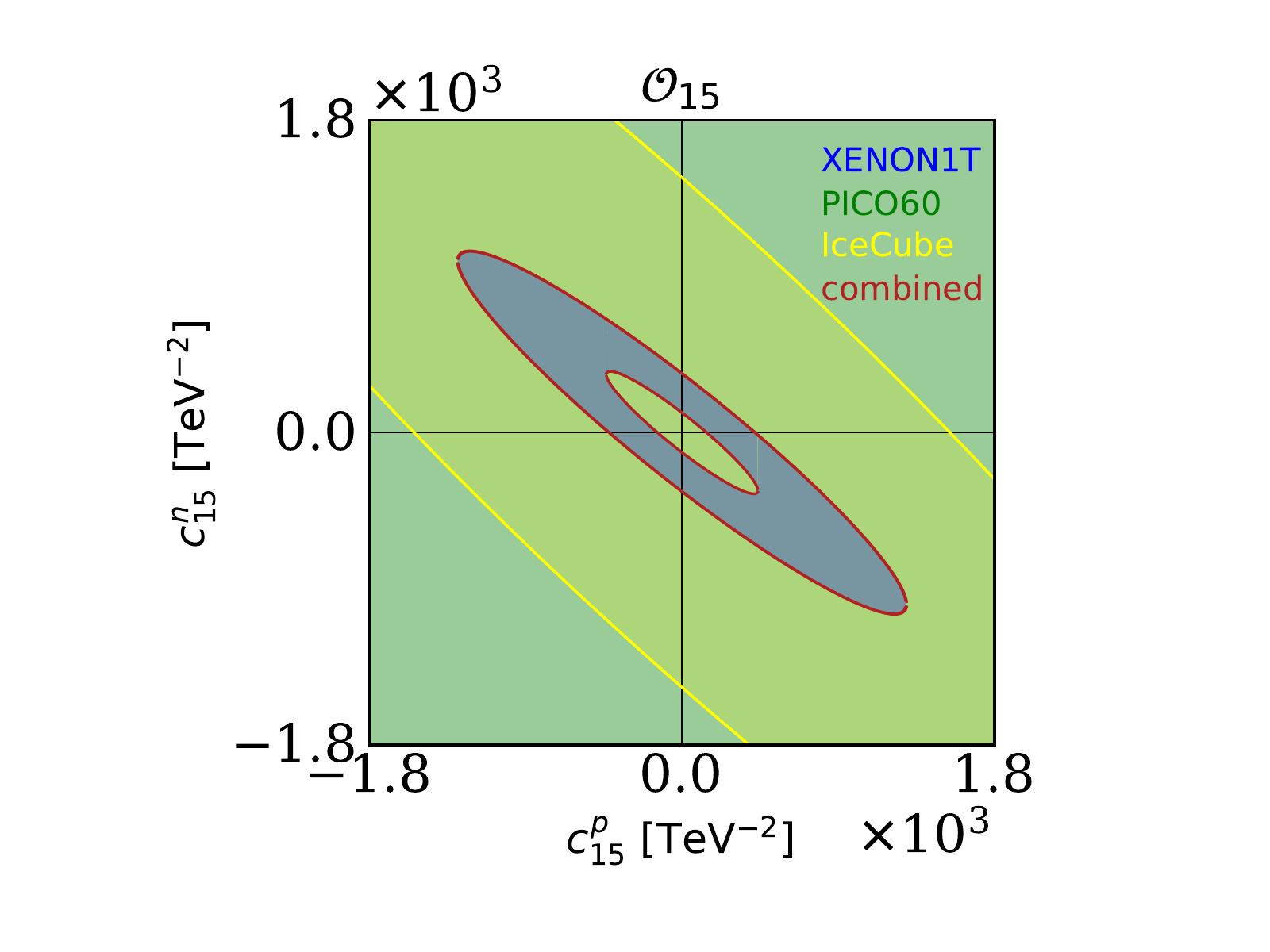}
		
	\end{center}
	\caption{90$\%$ C.L. allowed regions in the $c^p\times c^n$ parameter space from XENON1T (blue), PICO-60 (green), IceCube (yellow) and the combination of these (red), for $m_\chi=1$ TeV.}
	\label{fig:couplings_ellipses}
\end{figure}

The relaxation of the upper limits can be geometrically understood from Fig.~\ref{fig:couplings_ellipses}, which shows the 90\% C.L. allowed regions in the $c_i^p-c_i^n$ parameter space by the XENON1T (blue), PICO-60 (green) and IceCube (yellow) experiments, for each Galilean invariant operator $i=1, 3, ..., 15$. These plots are a rigorous version of the sketch presented in Fig.~\ref{fig:allowed_regions_demo}. One should note, however, that for the XENON1T experiment the allowed region does not include the point $c_i^n=c_i^p=0$. This is due to the large number of signal events observed by XENON1T compared to the background, and which may be interpreted as a hint for dark matter. In our work we will disregard this intriguing  possibility, and instead we will just focus on the derivation of model independent upper limits on the coupling strengths. 

The impact of the interference among the isoscalar and the isovector interactions in the limits on the coupling strengths to the nucleons is determined by the elongation and orientation of the allowed regions in the $c_i^p-c_i^n$ parameter space, and is most notable when the allowed region is very elongated and misaligned with the $c_i^p-c_i^n$ axes. The impact of the interference between different Galilean invariant operators can be sizable in some instances, such as in the limits from the XENON1T experiment on the ${\cal O}_{12}$ interaction, due to interference with the ${\cal O}_{11}$ and ${\cal O}_{15}$ interactions (see Appendix \ref{app:wimp_eft} for a description of all possible interferences among operators).

The interplay of such two effects can be in some cases the result of a delicate balance. For instance, in the $c_i^p-c_i^n$ base and for interactions depending on the $\Sigma^{\prime\prime}$ and $\Sigma^{\prime}$ nuclear response functions, the allowed regions are approximately aligned to the axes for both fluorine in PICO-60 and xenon in XENON1T (the former is a proton-odd nucleus where the spin contribution from neutrons almost cancels while the latter is neutron--odd, where the same happens for the spin contribution from protons). In general this should imply a mild relaxation of the bounds (defined as the ratio between the model independent bound on a coupling and that obtained by assuming that such coupling is the only non--vanishing one in the effective theory). However for the same interactions and in the case of PICO-60 the elongation of the allowed region can be very pronounced due to large cancellations among the interfering contributions to the expected rates from different couplings. In this case the effect from the very large elongation of the allowed region prevails over the approximate alignment to the axes, and the relaxation of the PICO-60 bounds turns out to be as large as $\sim$ 2 orders of magnitude.  

The specific choice of parameters that leads to such large cancellations in the calculation of the model independent bounds does not only lead to significantly weaker constraints: more importantly,  the robustness of the corresponding bounds becomes questionable both because one should go beyond the leading order in the calculation of the rate, and because at some stage the level of cancellation among the different $W_{Tk}^{\tau\tau^{\prime}}$ functions is expected to exceed the accurateness with which they are calculated (notice that the $W_{Tk}^{\tau\tau^{\prime}}$'s are numerical solutions of nuclear shell models that are usually approximated, as we do, with the polynomial fits provided in~\cite{haxton2,catena}). Moreover, such cancellations imply the presence of a very large hierarchy in the eigenvalues of the ensuing matrices, which can lead to numerical instabilities in the implementation of our method.~\footnote{The impact of the numerical instabilities in the calculation of the upper limits on the coupling strengths is discussed in Appendix \ref{app:error_propagation}.}
Going beyond the case of a single coupling and exploiting the interference among different operators may further weaken the bounds, although usually by no more than a factor of 2 compared to the isospin interference of the single coupling case. An exception to this pattern is the  ${\cal O}_8$--${\cal O}_9$ interference in fluorine, for which we observe a relaxation of the bound on either $c_i^p$ or $c_i^n$  close to two orders of magnitude, while at the level of isospin interference the same quantity for ${\cal O}_8$ alone is less than a factor of 2.

Fig. \ref{fig:allowed_regions_demo} also illustrates the importance of employing different targets to close in on the parameter space of spin-dependent DM-nucleon interactions. The blue cross is allowed by experiment ${\mathscr E}$, but lies well outside the allowed region of the experiment ${\mathscr E}^\prime$. An allowed point must necessarily lie in the intersection of the allowed regions by both experiments, however, clearly this is not sufficient, since points very close to the boundary of the intersection region could be excluded by more than 90$\%$ C.L. when combining the results of both experiments. In the next section we extend our method to provide a model independent upper limit on the coupling strength $c_\alpha$ from combining more than one experiment.

\section{Limits on the coupling strengths from combining several experiments}\label{sec:combined}

The formalism presented in Section~\ref{sec:single} can be extended to combine the results of $n$ experiments, and to provide a combined 90\% C.L. on a coupling strength. In this case, the Lagrangian Eq.~(\ref{eq:single_lagrangian}) must be replaced by
\be
	L\,=\,c_\alpha-\lambda \Big[ \chi_{\rm tot}^2({\bf c})-\chi_{{\rm tot}, \rm min}^2  \,-\, 2.71 \Big],
	\label{eq:comb_lagrangian}
\ee
where the total $\chi^2$ is defined as
\be
\chi_{\rm tot}^2({\bf c})= \sum_{\mathscr E}\chi_{{\mathscr E}}^2({\bf c})\;,
\ee
with the summation running over all experiments considered. Following the same steps as in Section~\ref{sec:single}, one obtains 
\begin{align}
	c_\beta^{\rm max}\, &=\frac{1}{2\lambda} (\mathbb{X}^{-1})_{\beta\alpha},
	\label{eq:cmax(gamma)_several}
\end{align}
where
\begin{align}
\mathbb{X}=\sum_{\mathscr E}
\Big[2 a_{\mathscr E}N_{\mathscr E}^{\rm sig}({\bf c}^{\rm max})+b_{\mathscr E}\Big]
\mathbb{N}_{\mathscr E}\;.
\label{eq:combinedmatrix}
\end{align}
Substituting in Eq.~(\ref{eq:Nsignal}) for each experiment one obtains $n$  implicit equations for the number of events at the $n$ experiments, $N_{\mathscr E}^{\rm sig}({\bf c}^{\rm max})$, of the form: 
\begin{align}
N_{\mathscr E}^{\rm sig}({\bf c}^{\rm max})=\frac{1}{4\lambda^2}
(\mathbb{X}^{-1} 
\mathbb{N}_{\mathscr E} \mathbb{X}^{-1})_{\alpha\alpha}\;.
\label{eq:Nsig_manyE}
\end{align}
These $n$ equations, along with the requirement 
\be
\chi_{\rm tot}^2({\bf c})-\chi_{{\rm tot}, \rm min}^2  = 2.71 ,
	\label{eq:comb_upperlimit}
\ee
lead to a solution for $N_{\mathscr E}^{\rm sig}({\bf c}^{\rm max})$ and $\lambda$. Finally, from Eq.~(\ref{eq:cmax(gamma)_several}) one obtains the values of the coupling strengths $c_\beta^{\rm max}$ at the point in parameter space that maximizes the coordinate $c_\alpha$, and specifically 
\begin{align}
	c_\alpha^{\rm max}\, &=\frac{1}{2\lambda} (\mathbb{X}^{-1})_{\alpha\alpha}.
	\label{eq:cmax(alpha)_several}
\end{align}

\begin{figure}[t!]
\begin{center}
\includegraphics[trim=120 200 120 300,clip,width=0.70\textwidth]{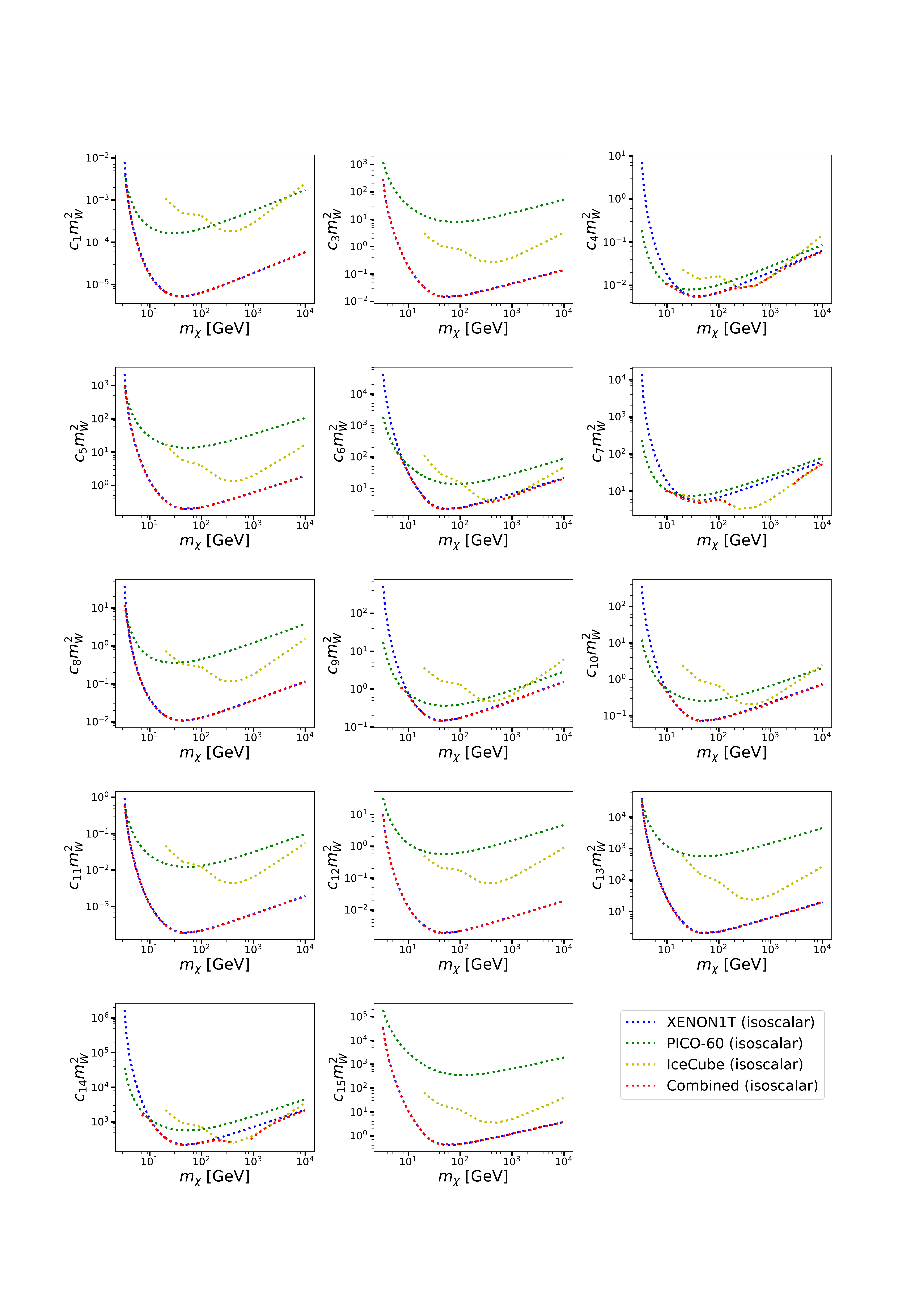}
\end{center}
\caption{90\%  C.L upper limits on the  dark matter-nucleon coupling from XENON1T, PICO-60 and IceCube, assuming only the isoscalar interaction. The dotted red line shows the limit from combining the three experiments. } 
\label{fig:isoscalar}
\end{figure}

We show in Fig.~\ref{fig:isoscalar} the upper limits on the DM-nucleon coupling strengths assuming the isoscalar interaction only, from combining the results from XENON1T, PICO-60 and IceCube. We also show for comparison the limits from each individual experiment. In this case, the combined limit for a given dark matter mass is not significantly different to the limit of the most constraining experiment. One should note, however, that for some points of the parameter space the combined isoscalar limit does not exist. This is again due to the large number of events observed at the XENON1T experiment, and which in our simplified analysis cannot be reconciled with the null searches from PICO-60 and IceCube assuming the isoscalar interaction only. Correspondingly, the requirement $\chi_{\rm tot}^2-\chi^2_{\rm tot, min} \leq 2.71$ cannot be fulfilled with the isoscalar coupling only, and can only be fulfilled in the presence of isospin violation.

\begin{figure}[t]
\begin{center} 
\includegraphics[width=0.9\textwidth]{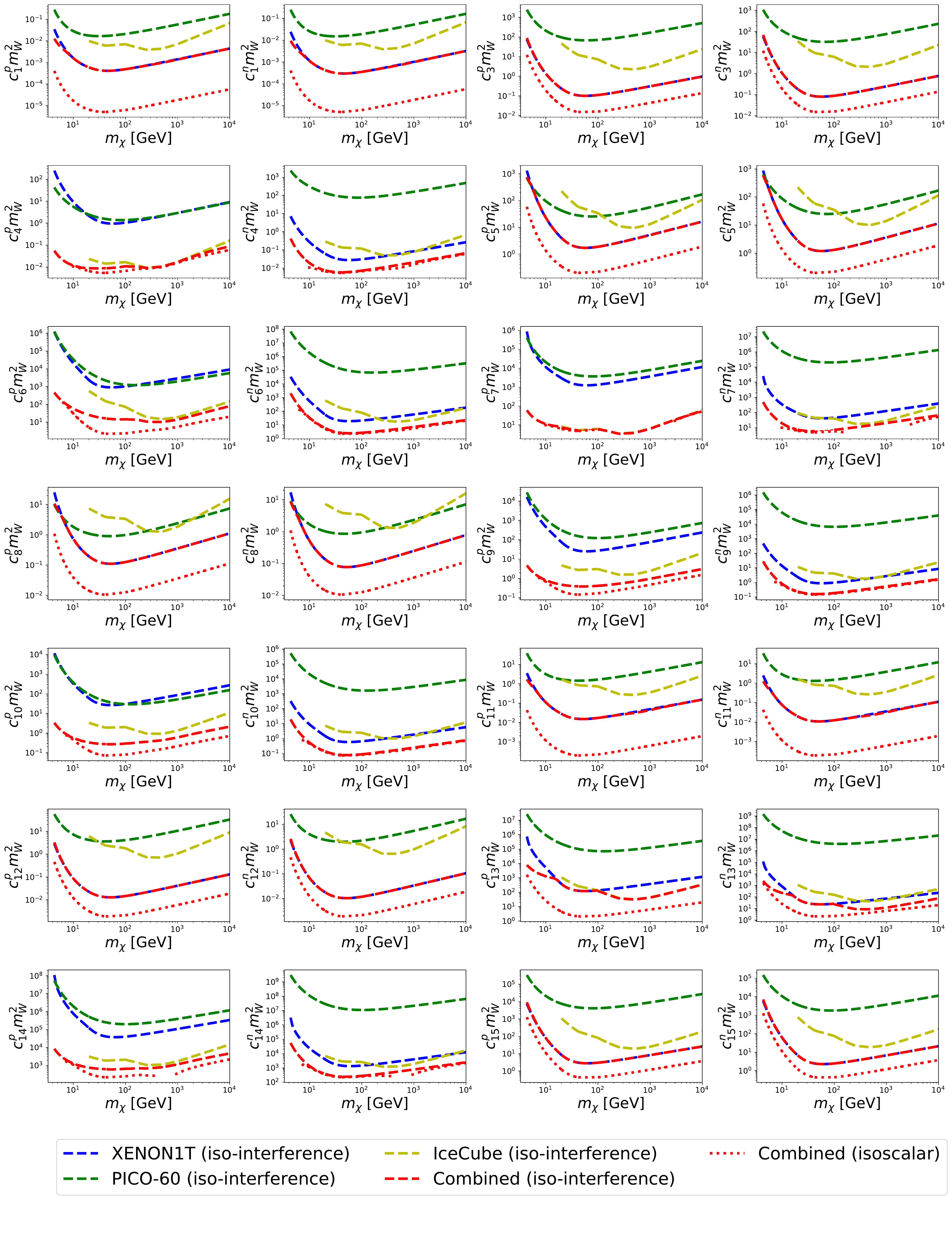}
\end{center}
\caption{Same as Fig.~\ref{fig:isoscalar}, but considering for each Galilean invariant operator the interference between the isoscalar and the isovector interaction. We also show for comparison the limit obtained assuming only the isoscalar interaction.}
\label{fig:isoin_comb_isos}
\end{figure}

The complementarity of experiments in probing the parameter space of the NREFT is more pronounced when including the interference among the isoscalar and isovector interactions. We show in  Fig.~\ref{fig:isoin_comb_isos} as a dashed red line the limits on the effective DM-proton (DM-neutron) coupling strengths $c^p_i$ ($c^n_i$) for $i = 1, 3, .., 15$ as a function of the dark matter mass, assuming that the isoscalar and the isovector interactions can interfere for a given operator ${\cal O}_i$. For some interactions, combining two experiments leads to an improvement of the iso-interference limits from single experiments, most notably for the operators ${\cal O}_4$, ${\cal O}_6$, ${\cal O}_7$, ${\cal O}_9$, ${\cal O}_{10}$, ${\cal O}_{13}$ and ${\cal O}_{14}$. For these interactions, the allowed regions are ``orthogonal" in the $c_i^n-c_i^p$ parameter space, as shown in Fig.~\ref{fig:couplings_ellipses}. In this case the scattering rate is driven by one of the two spin-dependent response functions $W_{\Sigma^{\prime\prime}}^{\tau\tau^{\prime}}$ or $W_{\Sigma^{\prime}}^{\tau\tau^{\prime}}$, and the complementarity between XENON1T, PICO-60 and IceCube is due to the fact that xenon has one unpaired neutron while fluorine and hydrogen have one unpaired proton.  For the remaining operators ${\cal O}_1$, ${\cal O}_3$, ${\cal O}_5$, ${\cal O}_8$, ${\cal O}_{11}$, ${\cal O}_{12}$ and ${\cal O}_{15}$ , the XENON1T allowed region lies completely within those from PICO-60 and IceCube for most values of the dark matter mass. Therefore, the combined iso-interference limit practically coincides with the XENON1T limit, except at low dark matter masses m$_{\chi} < 7$ GeV, where the sensitivity of PICO-60 is comparable to XENON1T. The non-complementarity of experiments for these operators can be explained from the nuclear response functions $W_{M}^{\tau\tau^{\prime}}$ or $W_{\Phi^{\prime\prime}}^{\tau\tau^{\prime}}$, that have a large hierarchy between xenon and fluorine/carbon/hydrogen/nitrogen. For comparison, we also show in the plot, as a dotted line, the combined limits derived assuming only the isoscalar interaction ({\it cf.} in Fig.~\ref{fig:isoscalar}).

\begin{figure}[t]
\begin{center}
\includegraphics[width=0.9\textwidth]{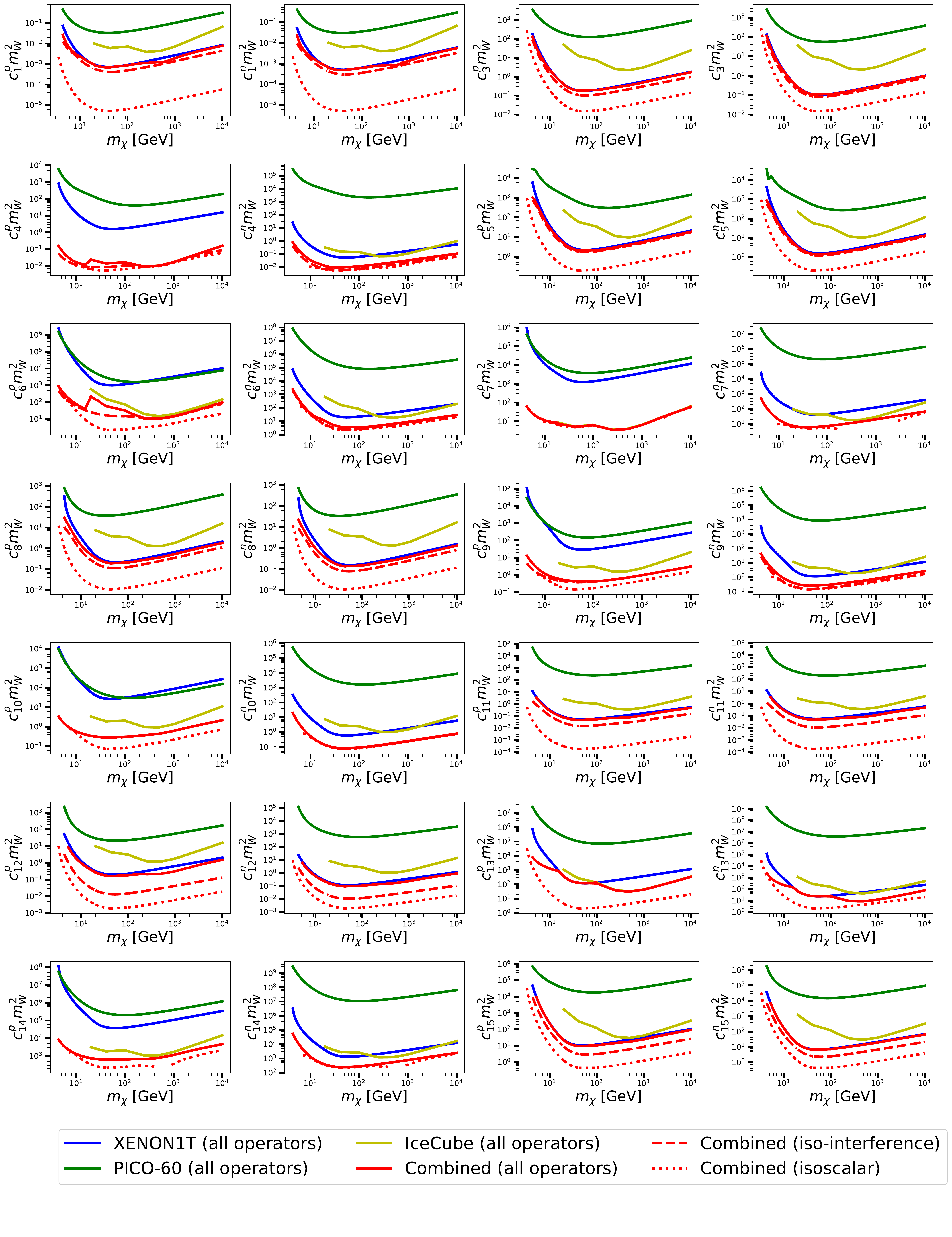}
\end{center}
\caption{Same as Fig.~\ref{fig:isoscalar}, but including the interference among different Galilean invariant operators. We also show for comparison the combined limits obtained assuming the isoscalar and iso-interference interactions.}
\label{fig:all_op_comb}
\end{figure}

\begin{figure}[t!]
\begin{center}
  \includegraphics[width=0.4\textwidth]{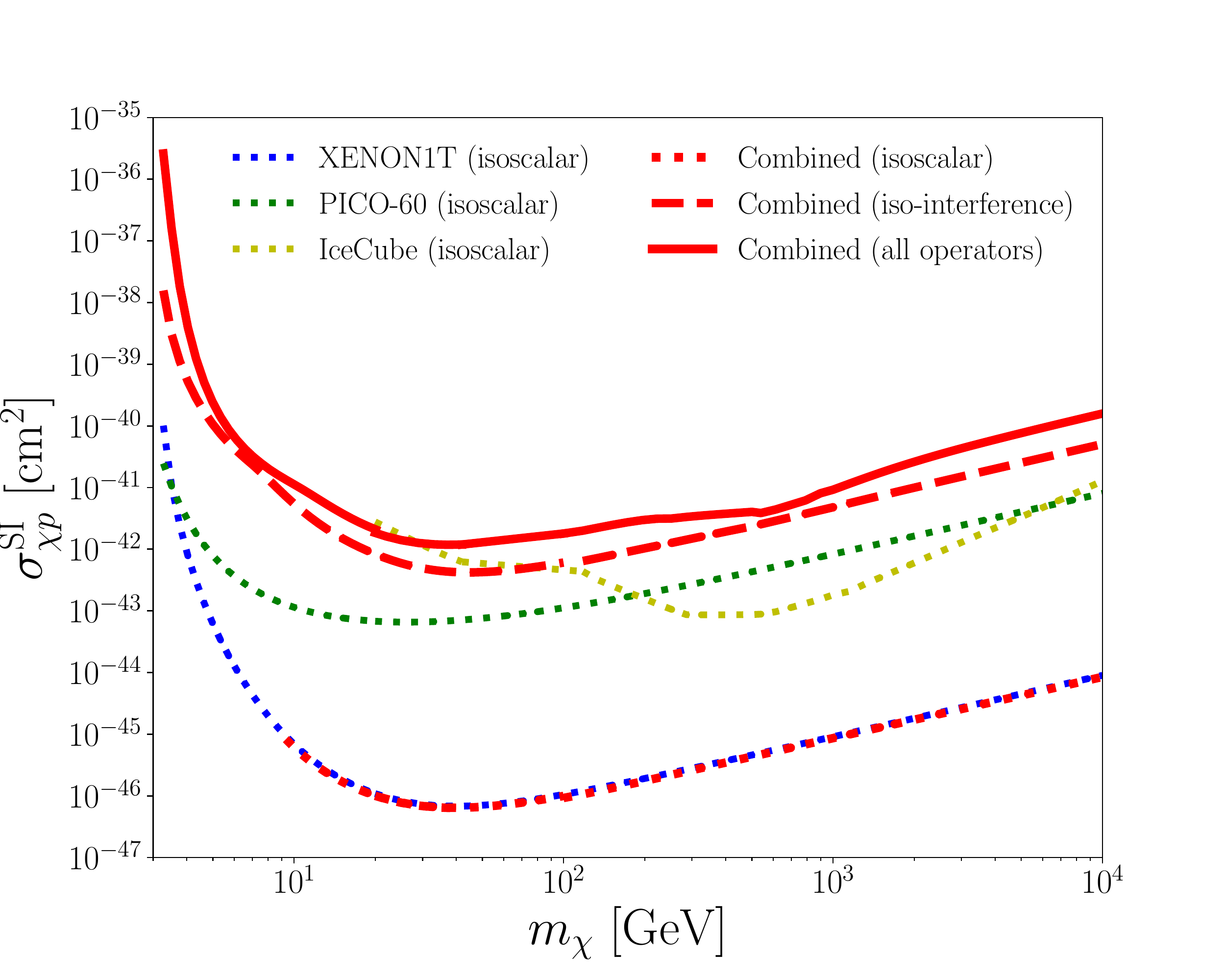}
  \includegraphics[width=0.4\textwidth]{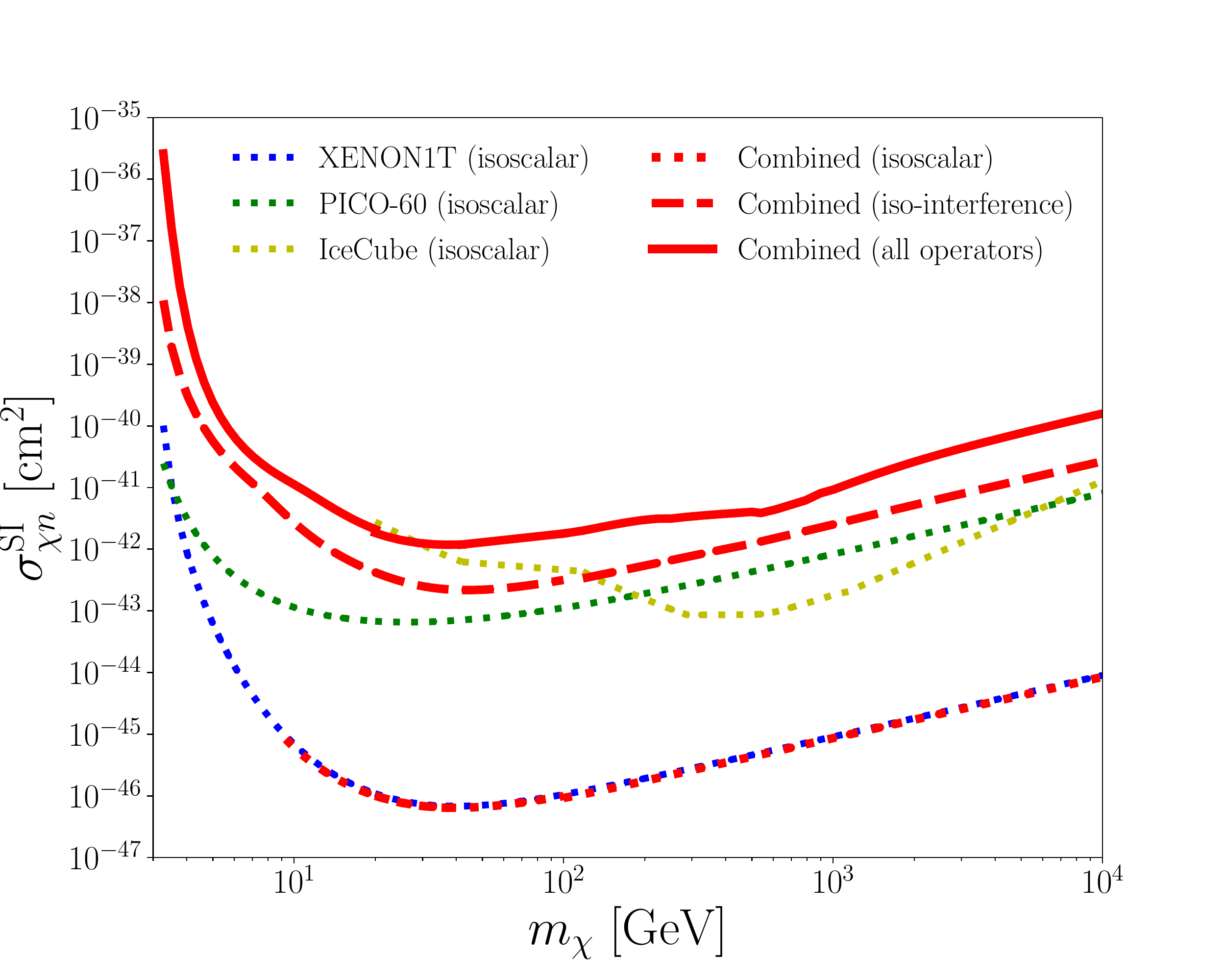}\\
  \includegraphics[width=0.4\textwidth]{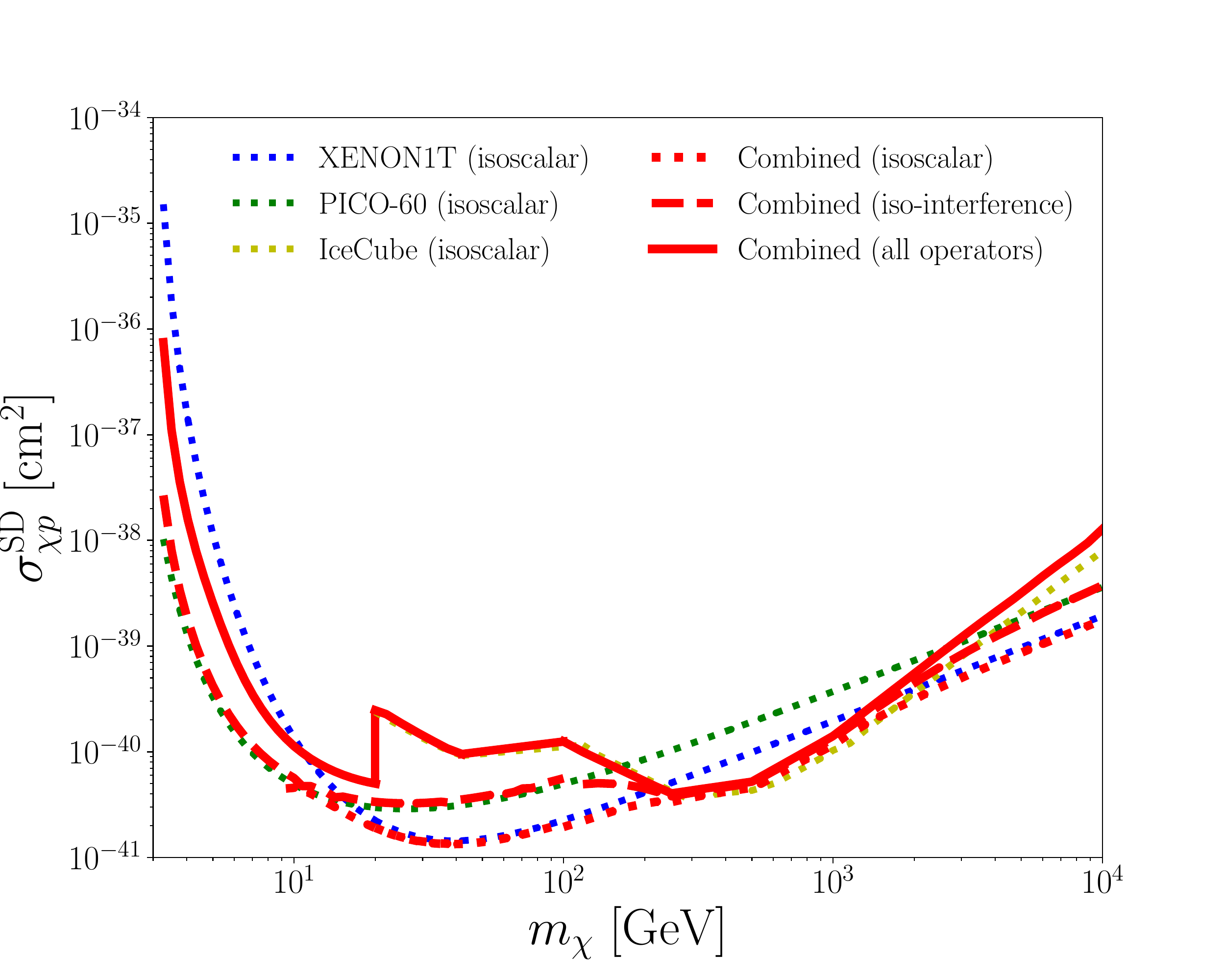}
  \includegraphics[width=0.4\textwidth]{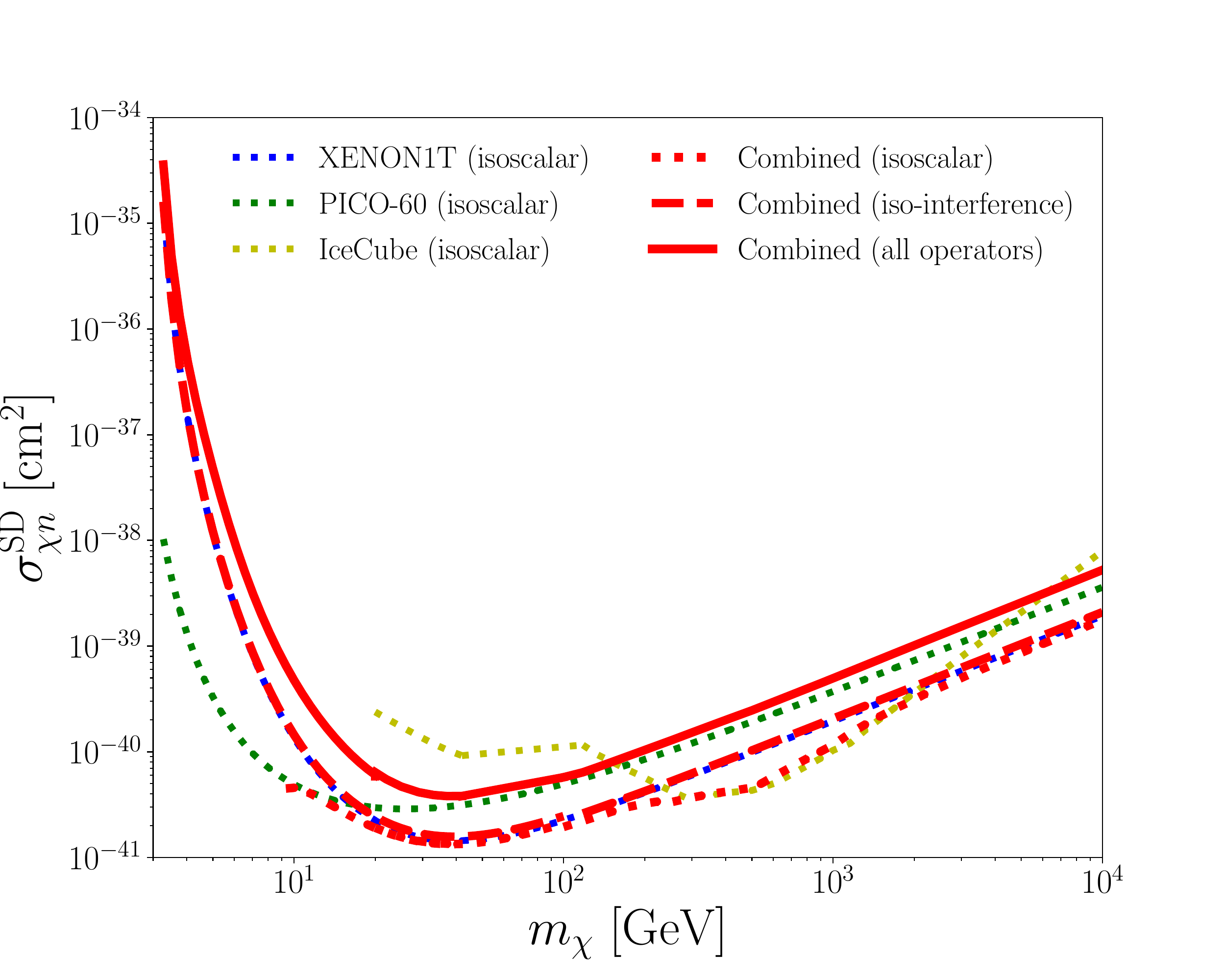}\\
\end{center}
\caption{Same as Fig.~\ref{fig:exclusions_SI_SD}, but showing the combined limit assuming only the isoscalar interaction for the given Galilean invariant operator (${\cal O}_1$ or ${\cal O}_4$ for SI or SD respectively, red dotted line), including the interference between the isoscalar and isovector interactions (red dashed line), or including the interference between all interactions and all Galilean invariant operators (red solid line). The plot also shows for comparison the isoscalar-only limits from XENON1T (blue), PICO-60 (green) and IceCube (yellow).}
  \label{fig:exclusions_comb_SI_SD}
\end{figure}

We show  in Fig.~\ref{fig:all_op_comb} as a solid red line the combined limits including the interference among Galilean invariant operators; the combined isoscalar and iso-interference limits are also shown, as red dotted and red dashed lines, for comparison. The interference among different operators further relaxes the upper limits on the coupling strengths, although the effect is small compared to the iso-interference. A notable exception is the  ${\cal O}_{12}$ operator, for which the interference with the ${\cal O}_{11}$ and ${\cal O}_{15}$ operators leads to limits which are about one order of magnitude weaker than those derived considering the ${\cal O}_{12}$ operator only.  Lastly, and in order to compare with the published limits, we show in Fig. \ref{fig:exclusions_comb_SI_SD} the limits for the coupling strengths of the operators ${\cal O}_1$ and ${\cal O}_4$ recast into spin-independent (top panels) and spin-dependent (bottom panels) DM-proton (left panels) and DM-neutron (right panels) interactions cross-sections, using Eq.~(\ref{eq:sdcs}). 

\section{Conclusions}
\label{sec:conclusion}

We have investigated the complementarity of experiments in probing the parameter space of the NREFT of dark matter-nucleon interactions, with emphasis on scenarios where more than one interaction is present, for instance, when the dark matter couples to the nucleus via both the isoscalar and isovector interaction, or when the effective theory of that scenario contains more than one Galilean invariant operator. To this end, we have developed a method to determine the 90\% C.L. upper limit of a given coupling strength of the NREFT from the non-observation of a signal in a set of experiments, including the interference of operators. The resulting limits can then be applied to any model, in contrast to the limits published by the experimental collaborations, which implicitly assume equal coupling to protons and to neutrons, and only the spin-independent or the spin-dependent interactions. The method can be extended to the case where one (or several) experiments detect a signal, in order to determine the allowed ranges of a given coupling strength without making assumptions on the underlying physics. 

We have first applied this method to derive 90$\%$ C.L. upper limits on the DM-proton and DM-neutron coupling strengths from the null search results of a single experiment, concretely from XENON1T, PICO-60 or IceCube (assuming equilibrium between capture and annihilation, and assuming for concreteness annihilation into $W^+W^-$ for $m_\chi>100$ GeV and $\tau^+\tau^-$ for $m_\chi<100$ GeV). We find that the impact of the interference among operators can be very significant in some cases. Namely, the model independent limits on the DM-proton or the DM-neutron coupling strengths can be relaxed by several orders of magnitude compared to the published limits, derived under the assumption that the interaction is purely isoscalar.  

We have also emphasized the importance of using different targets in probing the vast parameter space of the NREFT of dark matter-nucleon interactions. Concretely, we have calculated upper limits on the  DM-proton and DM-neutron coupling strengths from combining the results of XENON1T, PICO-60 and IceCube, including the interference between the isoscalar and isovector interactions, and including the interference among all Galilean operators of the NREFT. The impact is particularly significant for those operators depending on spin-dependent response functions. For these operators, a given experiment loses sensitivity for concrete combinations of the couplings. However, these ``blind directions'' depend on the target material, and could be efficiently probed in an experiment employing a different target. In some instances, combining different experiments could be a better strategy to close in on the parameter space of the NREFT than increasing the exposure of a single experiment.

\acknowledgments 
The work of A.I, G.T., G.H., and A.B. was supported by the Collaborative Research Center SFB1258 and by the Deutsche Forschungsgemeinschaft (DFG, German Research Foundation) under Germany's Excellence
Strategy - EXC-2094 - 390783311. The work of S.K. and S.S. was supported by the National Research Foundation of Korea (NRF) funded by the Ministry of Education through the Center for Quantum Space Time (CQUeST) with grant number 2020R1A6A1A03047877 and by the Ministry of Science and ICT with grant number 2019R1F1A1052231. G.H is grateful to Riccardo Catena and Andreas Rappelt for useful discussions and suggestions.

\appendix

\section{Experimental likelihoods and upper limits}
\label{app:likelihoods}

If the parameters of a model can only take non-negative values, it is standard in particle physics to follow the CLs method to set confidence intervals on such parameters \cite{Read:2002hq}. We will follow the CLs method to determine the experimentally allowed regions for the coupling strengths in the $c_{i}^{p}-c_{i}^{n}$ basis, based in the likelihood-ratio test
\be\label{eq:test}
    \Lambda(m_{\rm \chi}, N^{\rm sig}_{\mathscr E})=-2\, {\rm ln}\Big[\frac{\mathcal{L}(m_{\chi},N^{\rm sig}_{\mathscr E})}{\mathcal{L}_{min}(m_{\chi}, N^{\rm sig}_{\mathscr E})}\Big],
\ee
where $\mathcal{L}(m_{\chi}, N^{\rm sig}_{\mathscr E})$ is the experimental likelihood of the signal + background hypothesis and $\mathcal{L}_{min}$ denotes the minimum of the likelihood function after varying over all parameters. By means of Wilks theorem \cite{Wilks:1938dza}, $\Lambda(m_{\chi}, N^{\rm sig}_{\mathscr E})$ converges asymptotically to a $\chi^{2}$ distribution with the desired significance and degrees of freedom equal to the difference in dimensionality between $\mathcal{L}$ and $\mathcal{L}_{\rm min}$. In this work, we are considering 3 experiments: XENON1T \cite{xenon_1t}, PICO-60~\cite{pico60_2015,pico60_2019}, and IceCube \cite{Aartsen:2016zhm}. For the single upper limits of XENON1T, PICO-60 (first bin), IceCube and the combined limits from these experiments, the degrees of freedom of the $\chi^{2}$ is one and the 90$\%$ C.L upper limit condition reads
\be\label{eq:combined}
 2\, {\rm ln}[\mathcal{L}_{min}(m_{\chi}, N^{\rm sig}_{\mathscr E})]-2\, {\rm ln}[\mathcal{L}(m_{\chi}, N^{\rm sig}_{\mathscr E})] = 2.71.
 \ee
For the single limit of PICO-60 (second bin), since the collaboration observed no events, $\mathcal{L}_{min}$ equals zero for every dark matter mass and the degrees of freedom are two in this case, giving us the 90$\%$ C.L upper limit condition
 \be
 -2\, {\rm ln}[\mathcal{L}(m_{\chi},N^{\rm sig}_{\mathscr E})] = 4.6.
 \ee

The likelihoods of XENON1T and PICO-60 have several parameters and are complicated to reproduce, therefore we will simply consider a poissonian likelihood of the number of observed events $N^{\rm obs}_{\mathscr E}$ and predicted signal $N^{\rm sig}_{\mathscr E}$ and background $\rm N^{\rm bck}_{\mathscr E}$ events

\be
\mathcal{L}(N^{\rm obs}_{\mathscr E} | N^{\rm sig}_{\mathscr E}+N^{\rm bck}_{\mathscr E})=\frac{(N^{\rm sig}_{\mathscr E}+N^{\rm bck}_{\mathscr E})^{N^{\rm obs}_{\mathscr E}}}{N^{\rm obs}_{\mathscr E} !} e^{-(N^{\rm sig}_{\mathscr E}+N^{\rm bck}_{\mathscr E})}.
\ee
with  $N^{\rm obs}_{\mathscr E}$ and  $N^{\rm bck}_{\mathscr E}$ given in Table~\ref{tab:exp_N}.

\begin{table}[t]
\begin{center}
\begin{tabular}{|c|c|c|}
\hline 
experiment &  $ N^{\rm obs}_{\mathscr E}$ & $N^{\rm bck}_{\mathscr E}$ \\
\hline
XENON1T &  14 & 7.36  \\
PICO-60 (1st bin) & 3   & 1  \\
PICO-60 (2nd bin) &  0 &  0 \\
IceCube & 926 & 931 \\
DeepCore & 427 & 414 \\
\hline
\end{tabular}
\end{center}
\caption{Number of observed and background events for each of the experiments considered in this work.}
\label{tab:exp_N}
\end{table}

The combined likelihood of a given set of n experiments reads
\be
\mathcal{L}_{\rm tot}= \prod_{\mathscr E =1}^{n}\mathcal{L}_{\mathscr E},
\ee
and taking the natural logarithm of the likelihood and multiplying by -2 we get
\be
    -2 \, {\rm ln[\mathcal{L}_{tot}]}= 2 \sum_{\mathscr E=1}^{n} \Big [N^{\rm sig}_{\mathscr E}+N^{\rm bck}_{\mathscr E}-N^{\rm obs}_{\mathscr E}\, {\rm ln} (N^{\rm sig}_{\mathscr E}+N^{\rm bck}_{\mathscr E})+{\rm ln}(N^{\rm obs}_{\mathscr E} !)\Big ].
\ee
In Section \ref{sec:combined}, we are following the Lagrange multipliers method to obtain the largest allowed values of the coupling strenghts from a set of $n$ experiments. This requires solving a system of $n+1$ equations, which includes the upper limit condition given by the test statistic of Eq.~(\ref{eq:test}). In our case, the upper limit condition is given by Eq.~(\ref{eq:combined}). Instead of solving the system of equations numerically, keeping the logarithm of the number of signal events in an explicit form, we approximate the likelihood functions with quadratic polynomials on the number of signal events. This allows us to keep track of the number of solutions and retain some analytical insight about our optimization problem. The corresponding fitting functions are given in Table~\ref{tab:exp_N}, and are accurate to within 1$\%$ in the relevant range for $
N^{\rm sig}_{\mathscr E}$.


\section{Optimization of coupling strengths: general formalism} \label{app:general_formalism}

In full generality, the Poissonian likelihood is given by
\be
\mathcal{L} (N_{\mathscr E}^{\rm sig}({\bf c})\big)=\frac{(N_{\mathscr E}^{\rm sig}({\bf c})+ N_{\mathscr E}^{\rm bck})^{N_{\mathscr E}^{\rm obs}}}{N_{\mathscr E}^{\rm obs} !} e^{-( N_{\mathscr E}^{\rm sig}({\bf c})+N_{\mathscr E}^{\rm bck})}.
\ee
Therefore, the associated  $\chi^2$ distribution Eq.~(\ref{eq:chi21}) explicitly reads
\be
	 \chi_{\mathscr E}^2({\bf c}) = 2 \left[N_{\mathscr E}^{\rm sig}({\bf c}) \,+\, N_{\mathscr E}^{\rm bck}\,-\, N_{\mathscr E}^{\rm obs}\,{\rm ln}\left(N_{\mathscr E}^{\rm sig}({\bf c})\,+\,N_{\mathscr E}^{\rm bck}\right)\,+{\rm ln}\left(N_{\mathscr E}^{\rm obs}!\right)\right],
	\label{eq:app_comb_chi2}
\ee
which has minimum
\be
 \chi_{\mathscr E,\rm min}^2 = 2 \left[N_{\mathscr E}^{\rm obs}-N_{\mathscr E}^{\rm obs}{\rm ln}
 (N_{\mathscr E}^{\rm obs})+{\rm ln}(N_{\mathscr E}^{\rm obs}!)
 \right].
 \label{eq:app_comb_chi2min}
\ee

Let us consider the general case with $n$ experiments. 
The optimization of the Lagrangian Eq.~(\ref{eq:comb_lagrangian}) leads in this case to the conditions 
\begin{align}
\frac{\partial L}{\partial c_\gamma}\Big|_{{\bf c}={\bf c}^{\rm max}} \,&=\,\delta_{ \beta \alpha}\,-\,4\lambda \sum_{\mathscr E}\left[ \,\left(1\,-\,\frac{N_{\mathscr E}^{\rm obs}}{N_{\mathscr E}^{\rm sig}({\bf c}^{\rm max})\,+\,N_{\mathscr E}^{\rm back}} \right) (\mathbb{N}_{\mathscr E})_{\beta \gamma} c_\gamma^{\rm max} \right] =\,0\;,
\label{eq:app_comb_max_zeta-equil}\\
\frac{\partial L}{\partial\lambda}\Big|_{{\bf c}={\bf c}^{\rm max}}\,&=
2 \sum_{\mathscr E} \left[N_{\mathscr E}^{\rm sig}({\bf c}^{\rm max})+N_{\mathscr E}^{\rm bck}-N_{\mathscr E}^{\rm obs}{\rm ln}\left(N_{\mathscr E}^{\rm sig}({\bf c}^{\rm max})+N_{\mathscr E}^{\rm bck}\right)+N^{\rm obs}_{\mathscr E}!\right] -\chi_{\mathscr E,{\rm min}}^2=2.71\;.
\label{eq:app_comb_max_lambda-equil}
\end{align}
From the first equation, one obtains an implicit equation for the $\beta$-th coordinate of ${\bf c}^{\rm max}$
\begin{align}
	c_\beta^{\rm max}\, &=\frac{1}{4\lambda}(\mathbb{X}^{-1})_{\beta\alpha},
	\label{eq:app_cmax(gamma)_app}
\end{align}
where $\mathbb{X}$ is a $28\times 28$ dimensional matrix defined as
\begin{align}
 \mathbb{X}\,&= \sum_{\mathscr E} \left(1\,-\,\frac{N_{\mathscr E}^{\rm obs}}{N_{\mathscr E}^{\rm sig}({\bf c}^{\rm max})\,+\,N_{\mathscr E}^{\rm bck}}\right)\mathbb{N}_{\mathscr E}  .
	\label{eq:app_xab}
\end{align}
Substituting this expression in Eq.~(\ref{eq:Nsignal}) one obtains an implicit equation for the number of signal events at the experiment  $\mathscr E$, in terms of the maximal number of events at every experiment considered and the Lagrange multiplier:
\begin{align}
N^{\rm sig}_{\mathscr E}({\bf c}^{\rm max})= \frac{1}{16\lambda^2}
\Big(\mathbb{X}^{-1}{\mathbb N}_{\mathscr E}\mathbb{X}^{-1}\Big)_{\alpha\alpha}.
\label{eq:app_Nsignalmax}
\end{align}
Using these $n$ equations and the requirement
\begin{align}
    \chi_{\rm tot}^2({\bf c})-\chi_{{\rm tot}, \rm min}^2 =2.71, 
\end{align}
with
\begin{align}
\chi_{\rm tot}^2({\bf c})= \sum_{\mathscr E}\chi_{{\mathscr E}}^2({\bf c}),
\end{align}
one can calculate the maximal number of events at the $n$ experiments and $\lambda$, and finally, the maximum possible value of the coupling strength $c_\alpha$ compatible at the 90\% C.L. with the $n$ experiments under consideration, and including the interference among operators
\begin{align}
	c_\alpha^{\rm max}\, &=\frac{1}{4\lambda}(\mathbb{X}^{-1})_{\alpha\alpha}.
\end{align}


\section{DM--nucleus cross section and detector response}
\label{app:wimp_eft}

In the present Appendix we give more details on the experimental matrices $\mathbb{N}_{\mathscr E}$ present in Eq. (\ref{eq:Nsignal}) and used in our analysis. In particular, we summarize the expressions for the DM--nucleus cross-section and we provide the formalism used in Section~\ref{sec:scattering_rate} to include the effect of the detector response.

In a direct detection experiment, the expected rate in the visible energy bin $E_1^{\prime}\le E^{\prime}\le E_2^{\prime}$ is given
by

\begin{eqnarray}
R_{[E_1^{\prime},E_2^{\prime}]}(t)&=&MT_{exp}\int_{E_1^{\prime}}^{E_2^{\prime}}\frac{dR}{d
  E^{\prime}}(t)\, dE^{\prime} \label{eq:start}\\
 \frac{dR}{d E^{\prime}}(t)&=&\sum_T \int_0^{\infty} \frac{dR_{\chi T}(t)}{dE_{ee}}{\cal
   G}_T(E^{\prime},E_{ee})\epsilon(E^{\prime})\label{eq:start2}\,d E_{ee} \\
E_{ee}&=&Q(E_R) E_R \label{eq:start3},
\end{eqnarray}

\noindent where $\epsilon(E^{\prime})\le 1$ is the experimental
efficiency/acceptance, $E_R$ (quoted in keVnr) is the recoil
energy deposited in the scattering process, and $E_{ee}$ (quoted in keVee) is the fraction of $E_R$ going to the experimentally detected process such as ionization, scintillation, and heat. While the factor $Q(E_R)$ represents the quenching factor, the symbol ${\cal
G_T}(E^{\prime},E_{ee}=q(E_R)E_R)$ denotes the probability of visible energy $E^{\prime}$ detection when DM scatters off an isotope $T$ in the detector target with recoil energy $E_R$. The fiducial mass and exposure of a detector are denoted by $M$ and $T_{exp}$ respectively. 

The differential rate, $\frac{d R_{\chi T}}{d E_R}$ is provided in Eq.~(\ref{eq:dr_de}), for which the differential cross-section is given by
\be
\frac{d\sigma_T}{d E_R}=\frac{2 m_T}{4\pi v_T^2}\left [ \frac{1}{2 j_{\chi}+1} \frac{1}{2 j_{T}+1}|\mathcal{M}_T|^2 \right ],
\label{eq:dsigma_de}
\ee
where the dark matter scattering amplitude on the target nucleus $T$ can be written in the
following form~\cite{haxton1,haxton2}

\be
  \frac{1}{2 j_{\chi}+1} \frac{1}{2 j_{T}+1}|\mathcal{M}|^2=
  \frac{4\pi}{2 j_{T}+1}
  \sum_{\tau=0,1}\sum_{\tau^{\prime}=0,1}\sum_{k}
  R_k^{\tau\tau^{\prime}}\left
  [c^{\tau}_i,c^{\tau^{\prime}}_i,(v^{\perp}_T)^2,\frac{q^2}{m_N^2}\right
  ] W_{T k}^{\tau\tau^{\prime}}(y).
\label{eq:squared_amplitude}
\ee

\noindent In the above expression $j_{\chi}$ and $j_{T}$ are dark matter
and the target nucleus spins, respectively, $q=|\vec{q}|$ while the
$R_k^{\tau\tau^{\prime}}$'s are DM response functions which depend on the couplings
$c^{\tau}_i$ as well as the transferred momentum $\vec{q}$ and
$(v^{\perp}_T)^2$. In Eq.~(\ref{eq:squared_amplitude}) the
$W^{\tau\tau^{\prime}}_{T k}(y)$'s are nuclear response functions and
the index $k$ represents different effective nuclear operators, which,
crucially, under the assumption that the nuclear ground state is an
approximate eigenstate of $P$ and $CP$, can be at most eight:
following the notation in \cite{haxton1,haxton2}, $k$=$M$,
$\Phi^{\prime\prime}$, $\Phi^{\prime\prime}M$,
$\tilde{\Phi}^{\prime}$, $\Sigma^{\prime\prime}$, $\Sigma^{\prime}$,
$\Delta$, $\Delta\Sigma^{\prime}$. The $W^{\tau\tau^{\prime}}_{T
  k}(y)$'s are function of $y\equiv (qb/2)^2$, where $b$ is the size
of the nucleus. For the target nuclei $T$ used in most direct
detection experiments the functions $W^{\tau\tau^{\prime}}_{T k}(y)$,
calculated using nuclear shell models, have been provided in
Refs.~\cite{haxton2,catena}.
In the decomposition form, the DM response function $R^{\tau\tau^{\prime}}_{k}$ is written as~\cite{haxton2},
\be
R_k^{\tau\tau^{\prime}}=R_{0k}^{\tau\tau^{\prime}}+R_{1k}^{\tau\tau^{\prime}}
(v^{\perp}_T)^2=R_{0k}^{\tau\tau^{\prime}}+R_{1k}^{\tau\tau^{\prime}}\left
(v_T^2-v_{min}^2\right ),
\label{eq:r_decomposition}
\ee

\noindent and its correspondence with the nuclear response functions is summarized in Table
\ref{table:eft_summary}. From the DM response functions one can appreciate that the following set of operators interfere: ${\cal O}_1$-${\cal O}_3$, ${\cal O}_4$-${\cal O}_5$-${\cal O}_6$, ${\cal O}_8$-${\cal O}_9$ and ${\cal O}_{11}$-${\cal O}_{12}$-${\cal O}_{15}$.   

\begin{table}[H]
\begin{center}
{\begin{tabular}{|c|c|c|c|c|c|}
\hline
\rule[-15.5pt]{0pt}{30pt}\multirow{2}{*}
{$\mathbf{c_j}$}  &  $R^{\tau \tau^{\prime}}_{0k}$  & $R^{\tau \tau^{\prime}}_{1k}$ & {$\mathbf{c_j}$}  &  $R^{\tau \tau^{\prime}}_{0k}$  & $R^{\tau \tau^{\prime}}_{1k}$ \\
\hline\hline
$c_1$  &   $M(q^0)$ & - & $c_3$  &   $\Phi^{\prime\prime}(q^4)$  & $\Sigma^{\prime}(q^2)$\\
$c_4$  & $\Sigma^{\prime\prime}(q^0)$,$\Sigma^{\prime}(q^0)$   & - & $c_5$  &   $\Delta(q^4)$  & $M(q^2)$\\
$c_6$  & $\Sigma^{\prime\prime}(q^4)$ & - & $c_7$  &  -  & $\Sigma^{\prime}(q^0)$\\
$c_8$  & $\Delta(q^2)$ & $M(q^0)$ & $c_9$  &  $\Sigma^{\prime}(q^2)$  & - \\
$c_{10}$  & $\Sigma^{\prime\prime}(q^2)$ & - & $c_{11}$  &  $M(q^2)$  & - \\
$c_{12}$  & $\Phi^{\prime\prime}(q^2)$,$\tilde{\Phi}^{\prime}(q^2)$ & $\Sigma^{\prime\prime}(q^0)$,$\Sigma^{\prime}(q^0)$ & $c_{13}$  & $\tilde{\Phi}^{\prime}(q^4)$  & $\Sigma^{\prime\prime}(q^2)$ \\
$c_{14}$  & - & $\Sigma^{\prime}(q^2)$ & $c_{15}$  & $\Phi^{\prime\prime}(q^6)$  & $\Sigma^{\prime}(q^4)$ \\
\hline
\end{tabular}}
\caption{Nuclear response functions corresponding to each coupling $c_i$ of the effective Hamiltonian (Eq.~\ref{eq:H}), for the velocity--independent and the velocity--dependent components of the DM response function, decomposed as in Eq.~(\ref{eq:r_decomposition}).  In parenthesis are the powers of $q$ in the DM response function.
\label{table:eft_summary}}
\end{center}
\end{table}


\section{Propagation of numerical errors}
\label{app:error_propagation}

The numerical calculation of the matrix $\mathbb{N}$ is unavoidably subject to errors, {\it e.g.} from the numerical algorithms of integration, from the limited precision in the calculation of the nuclear response functions, from the modeling of the nuclear effects or from the dark matter velocity distribution, to mention some. In this appendix we discuss how these errors would propagate in our calculation of the upper limits on the coupling strengths. 

Let us denote the true matrix as $\mathbb{N}_{\rm true}$, and the matrix calculated numerically as  $\mathbb{N}_{\rm num}$. The numerical matrix will slightly deviate from the true matrix by $\delta \mathbb{N}_{\rm true}$, which we assume a perturbation, so that $\mathbb{N}_{ \rm num}=\mathbb{N}_{ \rm true}+\delta \mathbb{N}$. For concreteness, let us assume that the matrix elements of the perturbation are of the form $(\delta \mathbb{N})_{ij}= \epsilon a_{ij} (\mathbb{N}_{\rm true})_{ij}$, namely that the relative error for all matrix elements is proportional to the small parameter $\epsilon\ll 1$. Here, $a_{ij}$ are ${\cal O}(0.1)$ parameters that take into account that the relative errors may depend on the matrix element. The elements of the true matrix are therefore related to the elements of the numerical matrix by
\begin{equation}
    (\mathbb{N}_{ \rm true})_{ij}=\frac{(\mathbb{N}_{ \rm num})_{ij}}{1+\epsilon\, a_{ij}}\simeq (\mathbb{N}_{ \rm num})_{ij} (1-\epsilon\, a_{ij})\;.
\end{equation}

The error in the determination of the matrix $\mathbb{N}$ will propagate to the calculation of the upper limits on the coupling strengths ({\rm cf.} Eq.~(\ref{eq:cmax_single})). The true upper limit on the coupling strength $c_\alpha$ is related to the numerical upper limit by
\begin{align}
(c^{\rm max}_{\alpha})_{\rm num}=
(c^{\rm max}_{\alpha})_{\rm true}
\sqrt{\frac{(\mathbb{N}^{-1}_{\rm num})_{\alpha\alpha}}{ (\mathbb{N}^{-1}_{ \rm true})_{\alpha\alpha}}}\;.
\end{align}
At first order in the perturbation, one obtains
\begin{align}
\mathbb{N}^{-1}_{\rm num}\simeq \mathbb{N}^{-1}_{\rm true}-\mathbb{N}^{-1}_{\rm true}\, \delta \mathbb{N}\, \mathbb{N}^{-1}_{\rm true}\;,
\end{align}
therefore
\begin{align}
(c^{\rm max}_{\alpha})_{\rm num}\simeq 
(c^{\rm max}_{\alpha})_{\rm true}
\Big[1-\frac{1}{2}\frac{\sum_{\beta\gamma}(\mathbb{N}^{-1}_{\rm true})_{\alpha\beta}(\delta\mathbb{N})_{\beta\gamma}(\mathbb{N}^{-1}_{\rm true})_{\gamma\alpha}}{ (\mathbb{N}^{-1}_{ \rm true})_{\alpha\alpha}}\Big]\;.
\label{eq:diff_num_true}
\end{align}

Using our ansatz for the perturbation, and the definition of the inverse of a matrix in terms of its adjugate and the determinant,  $\mathbb{N}^{-1}={\rm adj}({\mathbb{N}})/{\rm det}({\mathbb{N}})$, we obtain
\begin{align}
(c^{\rm max}_{\alpha})_{\rm num}\simeq 
(c^{\rm max}_{\alpha})_{\rm true}
\Big[1-\frac{1}{2}\frac{\epsilon}{{\rm det}(\mathbb{N}_{\rm true})}\frac{\sum_{\beta\gamma}
\Big({\rm adj}({\mathbb{N_{\rm true}}})\Big)_{\alpha\beta}
 a_{\beta\gamma} (\mathbb{N}_{\rm true})_{\beta\gamma}
\Big({\rm adj}({\mathbb{N_{\rm true}}})\Big)_{\gamma\alpha}}{ \Big({\rm adj}({\mathbb{N_{\rm true}}})\Big)_{\alpha\alpha}}\Big]\;.
\end{align}
In general, the correction term is small when $\epsilon\ll 1$ and the numerical result reproduces reasonably well the true result. However, in some special cases,  namely ${\rm det}(\mathbb{N}_{\rm true})$ is very small compared to the entries of $\mathbb{N}_{\rm true}$ and ${\rm adj}(\mathbb{N}_{\rm true})$ the correction can be sizable, or even lead to the breaking of the perturbative approach shown above. In the latter case, the value obtained numerically for $(c_\alpha^{\rm max})$ cannot be trusted, since it can be very different to the true value (or even be imaginary, if the elements of the diagonal of $\mathbb{N}_{\rm num}^{-1}$ turn out to be negative). This occurs when the eigenvalues of the matrix $(\mathbb{N}_{\rm true})$ are very hierarchical ({\it} i.e. when the allowed regions are very elongated). 

\begin{figure}[t!]
\begin{center}
    \includegraphics[width=0.4\textwidth]{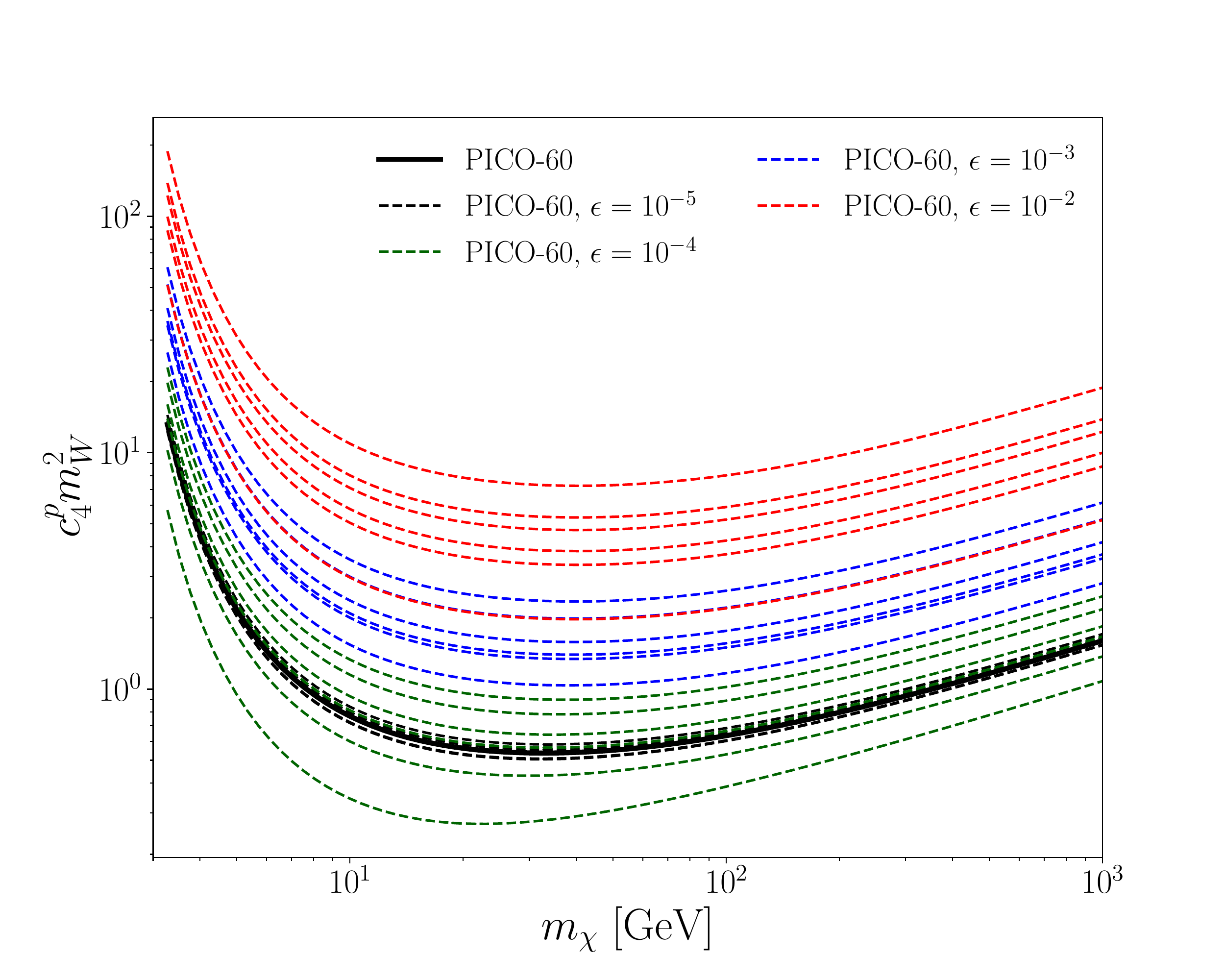}
  \includegraphics[width=0.4\textwidth]{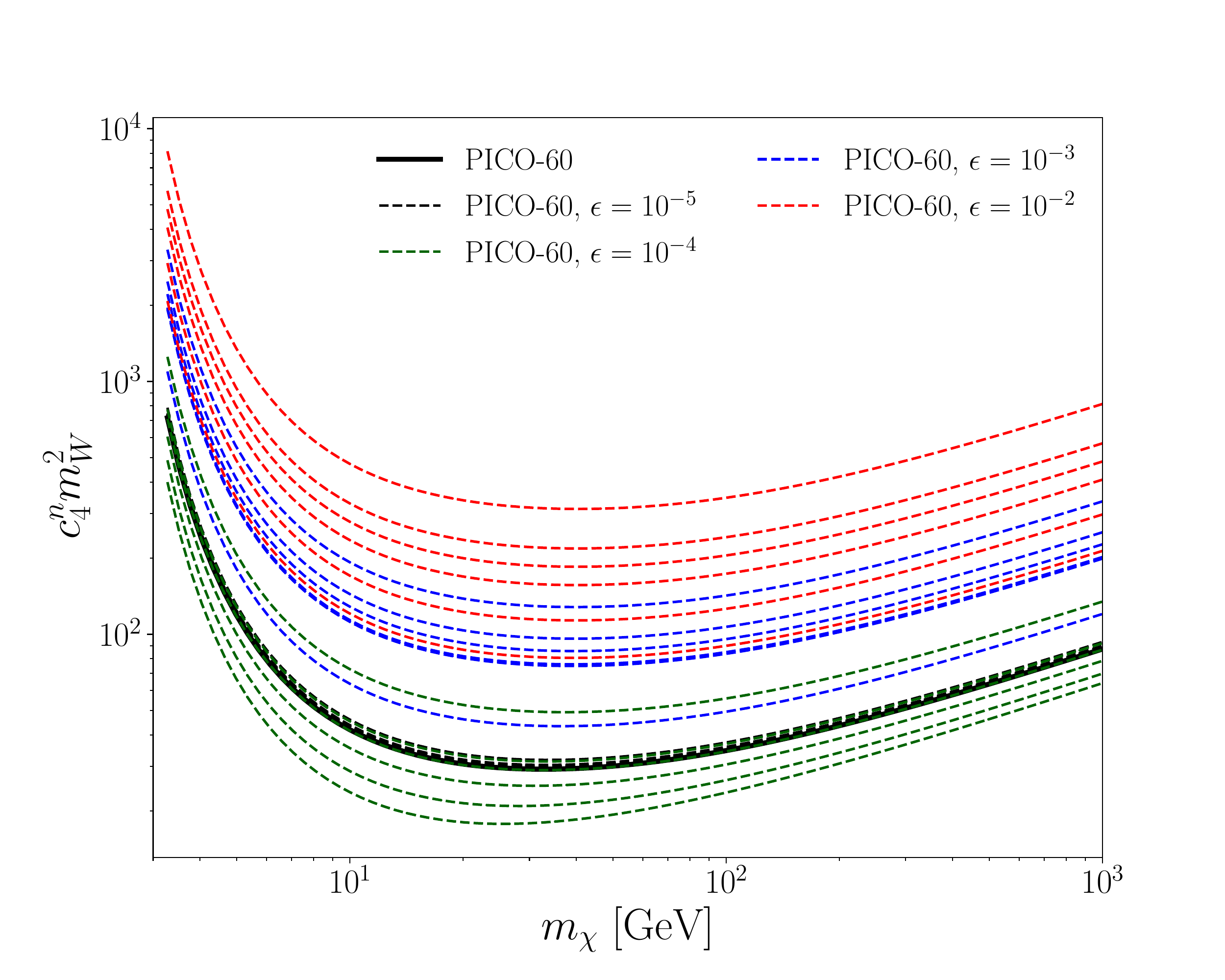}
\end{center}
\caption{Upper limits on the dark matter-proton (left plot) and dark matter-neutron (right plot) coupling strengths for the operator $\mathcal{O}_4$ from the PICO-60 experiment, including the interference between the isoscalar and isovector interactions, assuming that the relative error between the numerical and the true matrix elements in $\mathbb{N}_{\rm PICO-60}$ are of order  $\epsilon=10^{-5}, 10^{-4}, 10^{-3}, 10^{-2}$, for random values of $a_{ij}$ (see text for details).}
  \label{fig:o4epsilon}
\end{figure}

To be more specific, let us consider the simple case when $\mathbb{N}$ is a $2\times 2$ matrix, as occurs when calculating the limits on $c_i^p$ and  $c_i^n$ for the operator ${\cal O}_i$, including the interference between the isoscalar and isovector interactions. In this case, 
\begin{align}
(c_\alpha^{\rm max})_{\rm num}&\simeq 
(c_\alpha^{\rm max})_{\rm true}
\Big[1-\frac{\epsilon}{2}
\Big(a_{\alpha\alpha}+ (a_{11}+a_{22}-2a_{12})
\frac{(\mathbb{N}_{\rm true})_{12}^2}
{{\rm det}(\mathbb{N}_{\rm true})}\Big)\Big] \;,
\label{eq:two-dim}
\end{align}
for $c_1= c_i^n$ and $c_2= c_i^p$. From this expression, and for generic values of $a_{ij}$, it is clear that the upper limit can be trusted if 
\begin{align}
\epsilon\ll\frac
{{\rm det}(\mathbb{N}_{\rm true})}{(\mathbb{N}_{\rm true})_{12}^2}\;,
\end{align}
which is much more restrictive than the naive requirement $\epsilon\ll 1$. The impact of the propagation of errors in the upper limits on the coupling strengths is illustrated in Fig.~\ref{fig:o4epsilon}. This figure shows the upper limits on $c_4^p$ and $c_4^n$ from the PICO-60 experiment for different values of $\epsilon$. If the numerical precision of the calculation of the matrix elements of $\mathbb{N}_{\rm PICO-60}$ is better than $1\permil$, the calculated limit on $c_4^p$ differs from the true limit by an ${\cal O}(1)$ factor. However, if the precision is worse than 1\%, the numerical limit can differ from the true limit by more than one order of magnitude.

\bibliographystyle{JHEP} 

\providecommand{\href}[2]{#2}\begingroup\raggedright\endgroup

\end{document}